\documentclass{bmcart}

\usepackage{amsthm,amsmath}
\usepackage[utf8]{inputenc} 

\usepackage{graphicx}
\usepackage{amssymb}   
\usepackage{url}
\startlocaldefs
\newcommand{\lipe}{$^{6}$LiPE~}

\newcommand{\g}{$\gamma$}

\endlocaldefs

\begin{document}

\begin{frontmatter}

\begin{fmbox}
\dochead{Research}


\title{Simultaneous Gamma-Neutron Vision device: a portable and versatile tool for nuclear inspections}


\author[
  addressref={aff1},                   
  corref={aff1},                       
  email={jorge.lerendegui@ific.uv.es}   
]{\inits{J.}\fnm{Jorge} \snm{Lerendegui-Marco}}
\author[
  addressref={aff1},
]{\inits{V.}\fnm{V\'ictor} \snm{Babiano-Su{\'a}rez}}
\author[
  addressref={aff1},
]{\inits{J.}\fnm{Javier} \snm{Balibrea-Correa}}
\author[
  addressref={aff1},
]{\inits{L.}\fnm{Luis} \snm{Caballero}}
\author[
  addressref={aff1},
]{\inits{D.}\fnm{David} \snm{Calvo}}
\author[
  addressref={aff1},
]{\inits{I.}\fnm{Ion} \snm{Ladarescu}}
\author[
  addressref={aff1},
]{\inits{C.}\fnm{C\'esar} \snm{Domingo-Pardo}}


\address[id=aff1]{
  \orgname{Instituto de F{\'\i}sica Corpuscular (CSIC-University of Valencia)},          
  \city{Valencia},                              
  \cny{Spain}                                    
}



\end{fmbox}
\begin{abstractbox}
\begin{abstract} 
GN-Vision is a novel dual \g-ray and neutron imaging system, which aims at simultaneously obtaining information about the spatial origin of \g-ray and neutron sources. The proposed device is based on two position sensitive detection planes and exploits the Compton imaging technique for the imaging of \g-rays. In addition, spatial distributions of slow- and thermal-neutron sources ($<$100~eV) are reconstructed by using a passive neutron pin-hole collimator attached to the first detection plane. The proposed gamma-neutron imaging device could be of prime interest for nuclear safety and security applications. The two main advantages of this imaging system are its high efficiency and portability, making it well suited for nuclear applications were compactness and real-time imaging is important. This work presents the working principle and conceptual design of the GN-Vision system based on Monte Carlo simulations and explores its simultaneous \g-ray and neutron detection and imaging capabilities for a realistic case of a security inspection where a $^{252}$Cf source is hidden in a neutron moderating container. 
\end{abstract}

\begin{keyword}
\kwd{Gamma imaging}
\kwd{Neutron imaging}
\kwd{Nuclear inspections}
\kwd{Homeland security}
\kwd{Nuclear waste characterization}
\end{keyword}

\end{abstractbox}
%

\end{frontmatter}


\section{Introduction}\label{sec:Intro}
Simultaneous real-time imaging of \g-rays and neutrons is of interest for several nuclear safety and security applications such as control of reactor spent-fuel~\cite{Parker:15} and non-proliferation inspections of illicit production, use and trafficking of special nuclear material (SNM)~\cite{Poitrasson:15,Petrovic:21} or unmanned inspections in nuclear accidents~\cite{Sato:19,Vetter:18,Mochizuki:17}. In this context, existing systems with imaging capability for both \g-rays and neutrons are based on arrays of liquid scintillation detectors \cite{Pozzi:12,Madden:13,Poitrasson:14}, which are sensitive only to fast neutrons. However, in some situations the radioactive material is purposely attenuated or hidden by means of hydrogen-rich materials, thus leading to a thermal neutron spectrum. Moreover, fast neutron detectors present low intrinsic efficiencies and require large detection volumes. The latter represents a clear disadvantage in terms of portability and applicability. In this respect, several groups have been recently working on the development compact devices with dual neutron and \g-ray imaging capability~\cite{Hamrashdi:20,Steinberger:20}.

The use of dual neutron-gamma imaging devices has also potential interest in the field of hadron therapy. This methodology faces two important limitations related to real-time (neutron and gamma) dose monitoring~\cite{Schneider:15} and ion-beam range verification~\cite{Durante:19}, which limit the potential benefits of protons over photons. Dual neutron-gamma prototypes represent a promising approach to overcome these challenges but the size of most of the existing devices to date can also be a limitation for their implementation in clinical treatment rooms~\cite{Clarke:16}. 

In this context we present a new dual neutron- and \g-ray-imaging tool~\cite{Patent}, hereafter referred as GN-Vision, that aims at addressing the most relevant challenges for the aforementioned applications. In the present work we focus on its description and its potential application for identification of SNM. The system consists of a compact and handheld-portable device capable of measuring and simultaneously imaging both thermal- and slow-neutrons and \g-rays, both of them with a high efficiency. The proposed device consists of an upgrade of the i-TED Compton imager~\cite{Domingo16,Babiano20,Babiano:21,Lerendegui:22} developed within the ERC project HYMNS~\cite{hymns}, in such a way that simultaneous imaging of both \g-rays and slow neutrons becomes feasible with the same device.

This work presents the working principle and conceptual design of the GN-Vision system based on Monte Carlo simulations and demonstrates its simultaneous \g-ray and neutron detection and imaging capabilities. In Sec.~\ref{sec:concept}) we introduce the working principle of the proposed device and its evolution from the i-TED detector. Sec.~\ref{sec:MCdesign} deals with the technical implementation the GN-Vision, studied on the basis of Monte Carlo simulations. The results that demonstrate the dual capability to image \g-rays and neutrons are presented in Sec.~\ref{sec:Results}. After optimization of the conceptual design, the results of a simulated inspection of a container hiding nuclear material are shown in Sec.~\ref{sec:NuclearIns}. Last, a summary of our results and the outlook for the development of a first proof-of-concept prototype are provided in Sec.~\ref{sec:Summary}.


\section{Working principle of GN-Vision}\label{sec:concept}
 Simultaneous \g-ray and neutron imaging systems should fulfill several aspects~\cite{Hamrashdi:2019}. First, their active detection materials have to be sensitive to both types of particles and able to discriminate them. In addition, position sensitive detectors or multiple layers of detector arrays are required to reconstruct the interaction positions. Last, the use of either electronic or passive collimation techniques is required to reconstruct the spatial origin of the incoming radiation.

 Most of the existing dual imaging devices combine neutron and Compton scattering techniques to detect fast neutrons and \g-rays using combinations of organic liquid scintillators and high efficiency \g-ray detectors~\cite{Madden:13,Poitrasson:15,Clarke:16}. The former are sensitive to fast neutrons and \g-rays, and are able to classify each detected pulse as either coming from a neutron or a \g-ray interaction via pulse shape discrimination~\cite{Guerrero:08,Giaz:16,Blasi:18}. 

\begin{figure}[htbp!]
\begin{center}
\includegraphics[width=0.7\columnwidth]{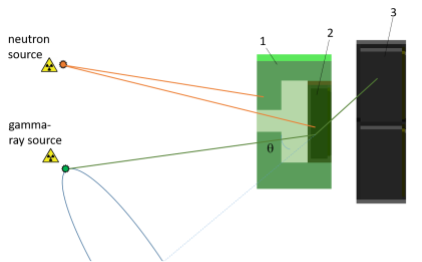}
\caption{Graphical representation of the working principle of the dual gamma-neutron imager GN-Vision comprising two position sensitive detection layers (2 and 3) and one passive neutron collimator (1).}
\label{fig:Concept}
\end{center}
\end{figure}

The proposed imaging device follows a novel working principle, sketched in Fig.~\ref{fig:Concept}:
\begin{itemize}
    \item The Compton imaging technique~\cite{Compton,Todd:74,Schonfelder:73} is exploited to detect and image \g-rays with energies of between 100 keV and several MeV using two detection planes, labelled (2) and (3) in Fig.~\ref{fig:Concept}. 
    
   \item The first detection plane (labelled (2) in Fig.~\ref{fig:Concept}) is able to detect neutrons of energies $<$1~keV and allows discriminating them from \g-rays. 
   
   \item  A passive neutron collimation system (labelled (1) in Fig.~\ref{fig:Concept}) attached to the first detection plane allows to carry out neutron imaging with the same working principle as pin-hole  cameras for \g-rays~\cite{Anger:58,Smeets:16,Caballero:18}.
\end{itemize}

For the imaging of \g-rays, our system operates as a Compton camera consisting of two position sensitive detection planes (see Fig.~\ref{fig:Concept}). The use of electronic collimation enhances significantly the detection efficiency when compared to passively collimated cameras and avoids additional structural material which will interact with neutrons as well~\cite{Domingo16}. The details of the Compton imaging technique can be found elsewhere~\cite{Compton,Domingo16}. In order to apply the Compton scattering law reliably, good resolution in both energy and position becomes mandatory~\cite{Babiano20}. This can be achieved by using scintillation crystals such as LaBr$_{3}$(Ce), LaCl$_{3}$(Ce) or CeBr$_{3}$, with high photon yield coupled to thin segmented photosensors, such as pixelated silicon Photomultipliers (SiPM). Aiming at enhancing detection efficiency, the two detection planes consist of large monolithic crystals. In the second detection layer the crystals are arranged in a compact configuration in order to cover a wide range of Compton angles ($\theta$)~\cite{Babiano20}.

 In order to achieve the imaging of neutrons, the active material of the first position sensitive detection layer of GN-Vision is chosen to have the capability of discriminating \g-rays and neutrons. A Cs$_{2}$LiYCl$_{6}$:Ce scintillation crystal enriched with $^{6}$Li at 95\% (CLYC-6), able of discriminating \g-rays, fast and thermal neutrons by Pulse Shape Discrimination (PSD), is a suitable option for this purpose~\cite{Giaz:16}. Slow neutrons reaching the first layer interact with CLYC-6 via the $^{6}$Li(n,$\alpha$)$^{3}$H reaction. The outgoing tritium and alpha particles deposit about 3.2 MeV in the crystal, which corresponds to an average penetration depth of only 54~$\mu$m and 13~$\mu$m, respectively, in contrast with the few cm range of Compton electrons for $\sim$MeV \g-rays. This means that, at variance with \g-ray imaging, the attainable position accuracy for thermal neutron imaging is remarkably higher than with high-energy \g-rays. As shown in Fig.~\ref{fig:Concept}, a pin-hole collimator made of a material with high absorption power for slow neutrons, is attached to the first plane. Knowing the geometry of the collimator (focal distance, aperture), from the measured response of the position-sensitive CLYC-6 detector one can reconstruct the 2D neutron-image (see Sec.~\ref{sec:ResultsNeutrons}). The latter has to be made of a low Z material which becomes essentially transparent to \g-rays of energies beyond 500~keV, thereby not affecting the Compton imaging performance and enabling simultaneous thermal-neutron and \g-ray vision (see Sec.~\ref{sec:ResultsGammas}).

\section{Design of the detector and Monte Carlo simulations}\label{sec:MCdesign}

\subsection{Technical design: from i-TED to GN-Vision}\label{sec:design}

From the technical point of view, the proposed GN-Vision device is an evolution of the \g-ray Compton imager i-TED developed within the HYMNS-ERC project~\cite{hymns}. i-TED is an array of four individual Compton imaging modules~\cite{Domingo16}, each of them consisting of two position-sensitive detection layers based on large monolithic LaCl$_{3}$(Ce) crystals. This novel imaging system has been fully characterized and optimized in the recent years~\cite{Olleros18,Babiano19} and the first demonstrator has been already assembled~\cite{Babiano20} and used in neutron TOF experiments~\cite{Babiano:21}. Last, advanced position reconstruction and full-energy selection algorithms based on Machine Learning techniques~\cite{Balibrea:21,Lerendegui:22,Balibrea:22} have been developed for i-TED. The GN-Vision device will profit from the aforementioned developments for an excellent efficiency and image resolution.

\begin{figure}[htbp!]
\begin{center}
  \includegraphics[width=0.7\columnwidth]{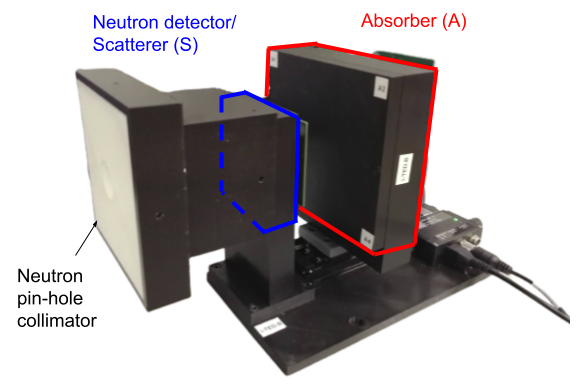} \end{center}
\caption{Possible mechanical implementation of the GN-Vision prototype where the three basic components of the device have been highlighted. The technical design is based on the previous i-TED Compton imager (see text for details).}
\label{fig:Technical}
\end{figure}

A possible design of the first GN-Vision prototype, based on the previous i-TED detector, is shown in Fig.~\ref{fig:Technical}. As mentioned in Sec.~\ref{sec:concept}, the first detection layer is made from a monolithic block of CLYC-6 scintillation crystal with a size of 50$\times$50$\times$10~mm$^3$, capable of fully absorbing neutrons below 100~eV. This detection plane will also act as the Compton scatterer detector (S) for the imaging of \g-rays (see Fig.~\ref{fig:Technical}). The \g-ray absorber detector (A) consists of an array of four LaCl$_{3}$ crystals, each one with a size of 50$\times$50$\times$25~mm$^3$. Each crystal base in both planes is coupled to a 2~mm thick quartz window, which is optically attached to a silicon photomultiplier (SiPM) from SensL (ArrayJ-60035-64P-PCB). The photosensor features 8$\times$8 pixels over a surface of 50$\times$50~mm$^2$ and is readout by means of front-end and processing PETsys TOFPET2 ASIC electronics~\cite{PETsys16}. The excellent time-response of these readout chips, originally developed for TOF-PET applications, enables one to implement them also for Compton imaging in a rather straight-forward and cost-effective manner~\cite{Babiano20,Babiano:21}.

In order to achieve the imaging of neutrons the first detection plane is supplemented with a neutron mechanical collimation system, as it is discussed in Sec.~\ref{sec:concept}. Among the low-Z neutron-absorbing materials, we chose highly (95\%) $^{6}$Li-enriched polyethylene ($^6$LiPE) due to its high absorbing power and simple mechanization. Moreover, $^{6}$Li among others since no \g-rays are emitted in the absorption of neutrons, thus avoiding a background source for the Compton imaging. For this collimation system a pin-hole approach has been considered in this work (see Figs.~\ref{fig:Technical} and \ref{fig:MC} ). A realistic design study of the proposed GN-Vision prototype, focusing on the critical parameters of the neutron collimator geometry, has been carried out by Monte-Carlo (MC) simulations, described in Sec.~\ref{sec:MC}.

\subsection{Monte Carlo simulations of GN-Vision}\label{sec:MC}

\begin{figure}[htbp!]
\begin{center}
  \includegraphics[width=0.7\columnwidth]{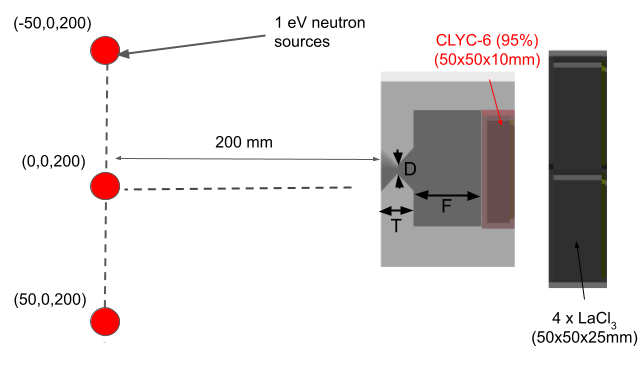}
 \end{center}
\caption{Schematic drawing of the geometry implemented in Geant4 for the proof-of-concept of GN-Vision. The main detector properties and the design parameters of the pin-hole collimator have been indicated (see text for details).}
\label{fig:MC}
\end{figure}

The design parameters of the GN-Vision system have been studied by means of MC simulations using the \textsc{Geant4} toolkit (v10.6)~\cite{Geant4_2}. This simulation study aimed at demonstrating the capability to image slow neutrons without affecting in a significant manner the \g-ray imaging efficiency and resolution. For this purpose, a detailed geometry model was implemented within Geant4 (see Fig.~\ref{fig:MC}, taking special care of the specifications for the composition of the CLYC-6 crystal~\cite{Giaz:16} and the $^{6}$LiPE~\cite{JohnCaunt}.

The modelling in \textsc{Geant4} of the physics processes can be carried out with different models, so-called Physics Lists (PL)~\cite{Geant4PL}. In this work, the simulations have been made using the officially released QGSP\_BIC\_HP Physics List~\cite{Geant4PL} which contains the standard Electromagnetic Package. For an accurate simulation of the neutron interactions, neutron-induced reactions below 20 MeV are simulated within this \textsc{Geant4} PL by means of the G4NeutronHP package~\cite{Mendoza:14}, using the G4NDL-4.6 data library (based on the JEFF-3.3~\cite{Plompen:20} evaluated data file).

The critical parameters in the design of GN-Vision, indicated in Fig.~\ref{fig:MC}, are the pinhole aperture (D), its thickness (T) and its focal distance (F). An additional parameter of relevance for the Compton imaging technique is the distance between the two detection planes. The latter establishes the balance between efficiency and angular resolution. Its impact in the detection efficiency is discussed in Sec.~\ref{sec:ResultsGammas}.
 
 To study the impact of the design parameters in the neutron imaging performance, three isotropic point-like sources of neutrons located at 20~cm from the collimator and separated from each other by 5~cm were simulated, as it is sketched in Fig.~\ref{fig:MC}. A total of of 10$^{8}$ neutrons were randomly generated from each of the sources. These simulations were carried out for different neutron energies ranging from thermal energies (25 meV) to 1 keV. Additional simulations of a point-like \g-ray source of energies ranging from 100 keV to 1 MeV and placed in the central position were carried out to study the impact of the neutron collimator in the detector response and the Compton images. The results for the optimization of the neutron imaging are presented in Sec.~\ref{sec:ResultsNeutrons} and the \g-ray imaging results are discussed in Sec.~\ref{sec:ResultsGammas}.

 The output of the MC simulation features the same format than the experimental data, including for each simulated event the deposited energy, interaction position and time of all the neutron and \g-ray hits in the two detection layers of GN-Vision. To mimic the discrimination of \g-rays and neutrons via PSD in the CLYC-6 crystal, a flag was included to identify energy depositions via $^{6}$Li(n,$\alpha$)$^{3}$H reactions from those carried out by electrons associated to \g-ray events. Experimental effects such as the low energy threshold, position and energy resolutions, which have been experimentally characterized for i-TED~\cite{Olleros18,Babiano19,Balibrea:21}, were included in the simulations to consider their impact on the imaging resolution.
 
\section{Results of the GN-Vision performance}\label{sec:Results}
\subsection{Neutron imaging}\label{sec:ResultsNeutrons}

The simple geometry of Fig.~\ref{fig:MC} served in this work to study the neutron imaging capabilities and optimize the design parameters of the pin-hole collimator for various neutron energies ranging from thermal (25.3~meV) up to 100~eV. 

In order to build the neutron images, we chose events in the CLYC-6 detector in which the energy deposition is carried out by an $\alpha$ particle and a triton. An additional cut in deposited energy around the 4.78~MeV peak, allows improving the selection and removing the contribution of fast neutrons. An energy window of $\pm$150~keV was applied to account for the expected energy resolution of about 3\% reported in previous works~\cite{Machrafi2014}. Once the slow-neutron events are selected, the image is reconstructed from the 3D-coordinates of the neutron hit in the CLYC-6 crystal by applying inversions in both the x and y planes and a scaling factor $S = d / f$, where $d$ is the distance from the collimator pin-hole to the plane where the neutron sources are placed, and $f$ is the distance of the pin-hole to the depth of interaction of the neutron. 

\begin{figure}[t!]
\begin{center}
  \includegraphics[width=0.45\columnwidth]{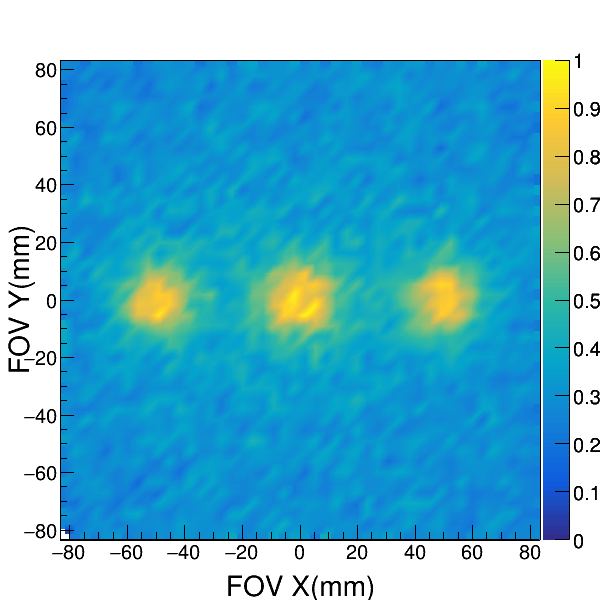}
  \includegraphics[width=0.45\columnwidth]{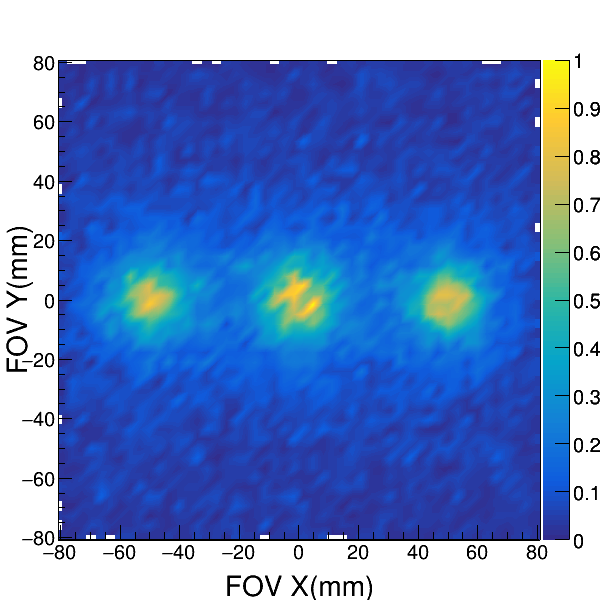}
 \end{center}
 
\caption{Reconstructed neutron images for three isotropic point-like sources of 1~eV neutrons located at 20 cm  from GN-Vision. This images have been obtained with a \lipe collimator has a focal F = 40~mm, a pin-hole diameter D = 2.5~mm and different thicknesses of T = 10 mm (left) and T = 30 mm (right). The images have been normalized to the maximum.}
\label{fig:NeutronImpactT}
\end{figure}

\begin{figure}[t!]
\begin{center}
  \includegraphics[width=0.45\columnwidth]{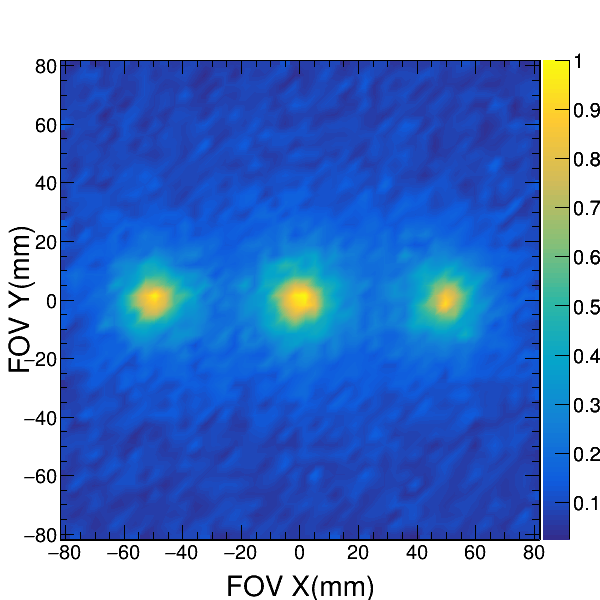}
  \includegraphics[width=0.45\columnwidth]{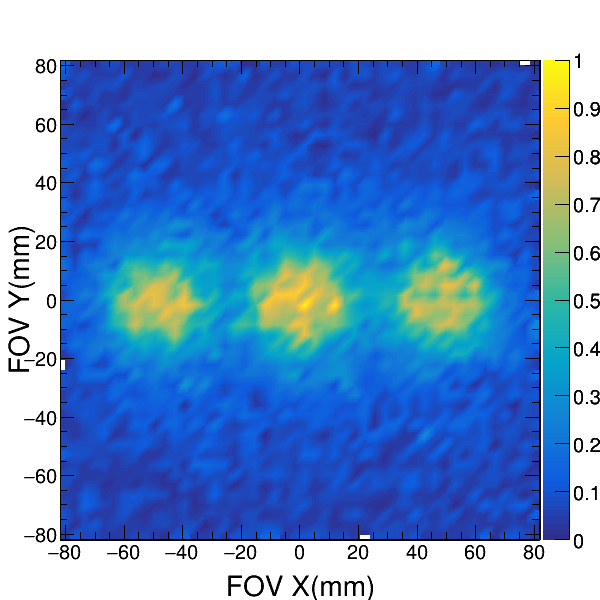}
 \end{center}
\caption{Reconstructed neutron images for three isotropic point-like sources of 1~eV neutrons located at 20 cm from GN-Vision. These images have been obtained using a \lipe collimator with a focal distance F = 40~mm, a thickness of T = 20 mm and a pin-hole diameter of D = 1 mm (left) and D = 5 mm (right). The images have been normalized to the maximum.}
\label{fig:NeutronImpactD}
\end{figure}

 Some examples of the reconstructed neutron images are presented in Figs.~\ref{fig:NeutronImpactT} and \ref{fig:NeutronImpactD},  proving the imaging capability of GN-Vision for slow neutrons. To illustrate the imaging capability the images corresponding to monoenergetic neutrons of 1~eV has been chosen as a representative energy in the range from thermal to 100~eV. From the reconstructed images, we have studied the role of two critical parameters of the neutron collimator: the collimator diameter (D) and its thickness (T). The remaining parameter, the focal distance F, only affects the size of the field of view and was adjusted to F=40~mm to image the three point-like sources of Fig.~\ref{fig:MC}. 

\begin{figure}[h!]
\begin{center}
  \includegraphics[width=0.45\columnwidth]{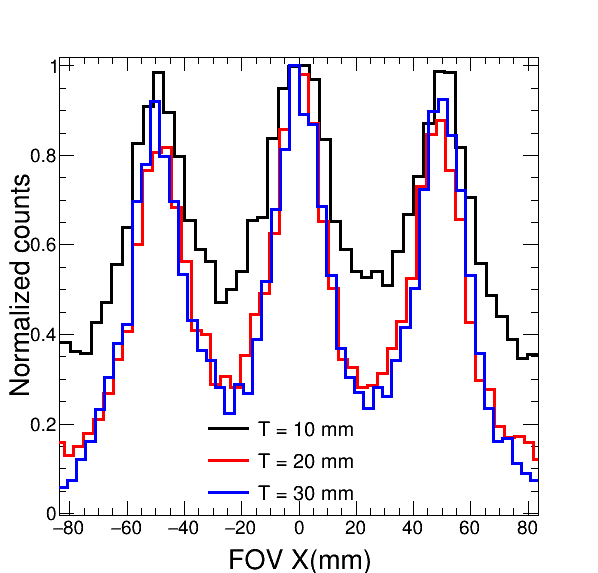}
  \includegraphics[width=0.45\columnwidth]{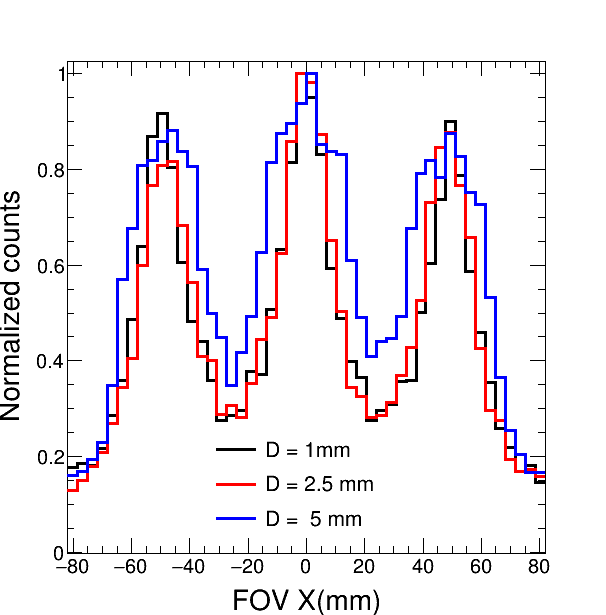}
 \end{center}
\caption{Normalized projections along the X-axis of the neutron images of Figs.~\ref{fig:NeutronImpactT} and \ref{fig:NeutronImpactD}. Left: impact of the collimator thickness (T) for a fixed diameter of D = 2.5 cm. Right: Impact of the collimator pin-hole diameter (D) for a fixed thickness of T = 20 mm. The focal is F = 40 mm for all the images.}
\label{fig:ProjectionsNeutronImages}
\end{figure}

Fig.~\ref{fig:NeutronImpactT} shows the role of the \lipe collimator thickness T, indicating than the contrast or peak-to-background ratio (PBR) of the image is enhanced with increasing thicknesses. A quantitative analysis can be done from the image projections, displayed in the left panel of Fig.~\ref{fig:ProjectionsNeutronImages}. From this figure we see that the PBR, calculated from the maximum divided by the value of the background underlying the images, increases from a factor 2.2 with T = 10 mm, to almost a factor of 10 with T = 30 mm.

\begin{table}[hb!]
\centering
\caption{Contrast of the neutron image (PBR) as a function of the neutron energy for various thicknesses (T) of the neutron absorbing collimator.}
\label{tab:Contrast}       
\begin{tabular}{cccc}
\hline
En (eV)& T = 10 mm & T = 20 mm   & T =40mm\\
 \hline
 0.025 &  14.73        &   18.95        & 46.16         \\
 1     &  2.23    &   5.86        & 14.05         \\
 10    &   1.41        &  2.31       &   7.56        \\
 100   &  1.21         &  1.31        &    2.42      \\
\hline
\end{tabular}
\end{table}

Aiming at a more comprehensive overview of the peformance of GN-Vision in the full neutron energy range under study, Table~\ref{tab:Contrast} summarizes the PBR obtained for the images of neutrons of different energies as a function of the neutron collimator thickness T. A thin \lipe of only 10~mm would be sufficient to achieve a PBR=15 for thermal neutrons while at least 40~mm of \lipe would be required to reconstruct an image with reasonable contrast, for instance PBR $>$2, for neutron energies beyond 100~eV.

The pin-hole diameter D has also a relevant impact on the reconstructed neutron images, as it can be seen in Fig.~\ref{fig:NeutronImpactD}. In this case, the attainable image resolution is clearly improved as the pin-hole size is reduced. The projections of the images obtained with different D values, shown in the right panel of Fig.~\ref{fig:ProjectionsNeutronImages}, illustrates that a reduction of the collimator from 5~mm to 1~mm improves the resolution (FWHM) of the images from 24~mm to 17~mm at the cost of a 50\% loss in detection efficiency. The interplay between these two magnitudes for the imaging of 1~eV neutrons is presented in Table~\ref{tab:Eff}. The values in this table correspond to the simulated setup of Fig.~\ref{fig:MC} with a collimator thickness T=20~mm.

\begin{table}[h!]
\centering
\caption{Absolute neutron imaging efficiency (neutron events/number of emitted neutrons) and imaging resolution (FWHM) of GN-Vision for 1~eV neutrons in the setup of Fig.~\ref{fig:MC}.}
\label{tab:Eff}       
\begin{tabular}{ccc}
\hline
Pin-hole diameter D (mm) & Efficiency &  Resolution (mm)\\
 \hline
1        &  3.1$\times$10$^{-5}$      & 17.4  \\
2.5      &  4.0$\times$10$^{-5}$         & 18.9 \\
5        &  6.0$\times$10$^{-5}$         & 24.1 \\
\hline
\end{tabular}
\end{table}

In this section, we have shown on the basis of Monte Carlo simulations that the proposed GN-Vision system is capable of imaging sources of low-energy neutrons. The first design of this device, based on attaching a simple neutron pin-hole collimator to the first detection plane, has led to the successful reconstruction of point-like neutron sources of energies below 100~eV with good spatial resolution and contrast. In terms of spatial resolution, the results of our device, with a resolution of 20~mm (FWHM), corresponding to 6$^{\circ}$, seems quite promising when compared to the 9$^\circ$-30$^{\circ}$ resolutions reported for other compact dual imaging systems sensitive to fast neutrons~\cite{Steinberger:20} and for large scintillator arrays~\cite{Poitrasson:14,Poitrasson:15}. Despite the high image resolution attainable with GN-Vision, the images reconstructed from slow neutrons in real scenarios are typically extended with respect to the true original source due to the moderation process around the emission point. Thus, for some cases and applications, this thermalization effect can provide also valuable information about the materials surrounding the neutron source, and their geometry, as it is reported later in Sec.~\ref{sec:NuclearIns}

The absolute efficiency values obtained for the imaging of low energy neutrons in this work (see Table~\ref{tab:Eff}) are close to the $10^{-4}$ reported for large liquid-scintillation arrays sensitive to fast neutrons~\cite{Poitrasson:14,Poitrasson:15}. The intrinsic neutron efficiency of the device is of the order of 2$\times$10$^{-3}$ for energies from thermal to 1~eV. Smaller efficiencies, ranging from 10$^{-3}$ and 10$^{-4}$ per incident neutron, have been reported for devices with comparable dimensions to GN-Vision~\cite{Hamrashdi:20,Steinberger:20}. The relatively large neutron imaging efficiency of the proposed device, which is directly related to the increasing neutron absorption cross section with decreasing neutron velocity and the intrinsically large thermal cross section of the $^{6}$Li(n,$\alpha$) reaction (940~barn), indicates the clear advantage of using slow neutrons for real-time imaging. The efficiency of the first GN-Vision design is still limited by the pin-hole geometry. More evolved designs, based for instance on coded-aperture masks~\cite{Cieslak:16} adapted to neutron detection, will be studied in the future to further enhance this key feature.

\subsection{\g-Ray imaging} \label{sec:ResultsGammas}

To demonstrate the dual imaging capability of GN-Vision we aim at showing in this section that the passive \lipe neutron collimator does not affect the imaging of \g-rays, which is accomplished by means of the Compton technique. 

\begin{figure}[t!]
\begin{center}
  \includegraphics[width=0.6\columnwidth]{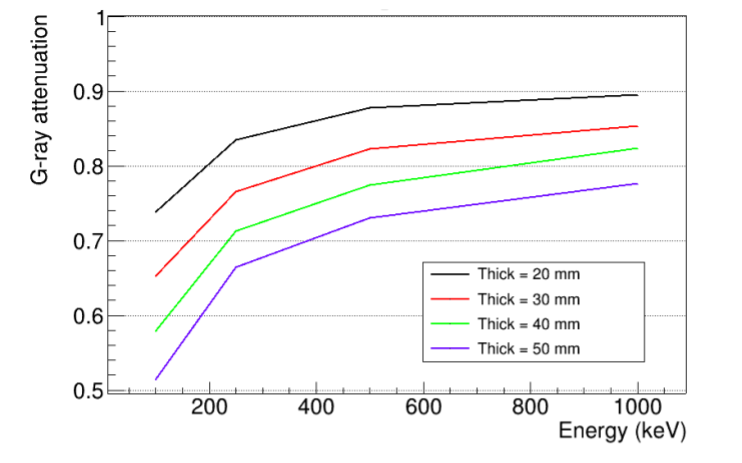}
 \end{center}
\caption{\g-ray attenuation factor related to the neutron collimator as a function of the energy for various thicknesses of the \lipe layer (T).}
\label{fig:GammaAttenuation}
\end{figure}

As introduced in Sec.~\ref{sec:MC}, the impact of the neutron collimator has been studied for mono-energetic \g-ray sources with energies below 1~MeV. Fig.~\ref{fig:GammaAttenuation} shows the ratio between the \g-ray events registered in the scatter detector as a function of the \g-ray energy for the thickness values of the \lipe collimator studied in Sec.~\ref{sec:ResultsNeutrons}. The neutron collimator of GN-Vision would absorb a fraction of the incoming \g-rays which ranges from 40\% for T = 40~mm and energies of 100~keV to just 10\% for T = 20~mm and 1~MeV  \g-rays, thus affecting the Compton imaging efficiency in the same proportion.

From the aforementioned simulations of mono-energetic \g-rays emitted from the central position of Fig.~\ref{fig:MC}, we have studied the impact of the neutron collimator in the Compton image. For the reconstruction of images via the Compton technique, only events in time coincidence between the Scatter (S) and Absorber (A) position-sensitive detectors (see Fig.~\ref{fig:Technical}) are considered from the output of the MC simulations. The deposited energies in the S- and A-layers together with the 3D-localisation of the \g-ray hits were extracted for each coincidence event. From these quantities, one can trace a cone, whose central axis corresponds to the straight line defined by the \g-ray interaction position in the two layers and its aperture $\theta$ is obtained from the measured energies using the Compton scattering formula (see for instance Eq.(2) of Ref.~\cite{Domingo16}). More details on the Compton technique and the implementation of the imaging algorithms can be found in previous works of the predecessor i-TED detector~\cite{Babiano20,Babiano:21,Lerendegui:22}. 

\begin{figure}[t!]
\begin{center}
  \includegraphics[width=0.9\columnwidth]{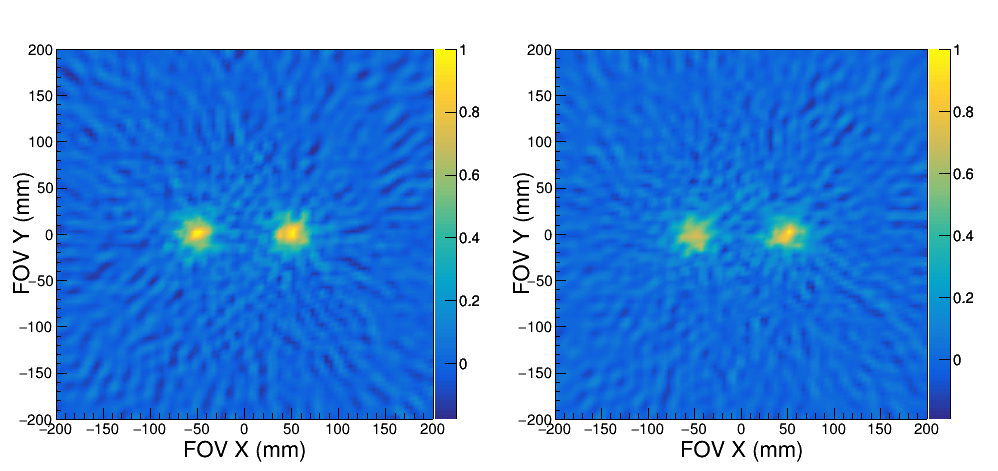}  \includegraphics[width=0.45\columnwidth]{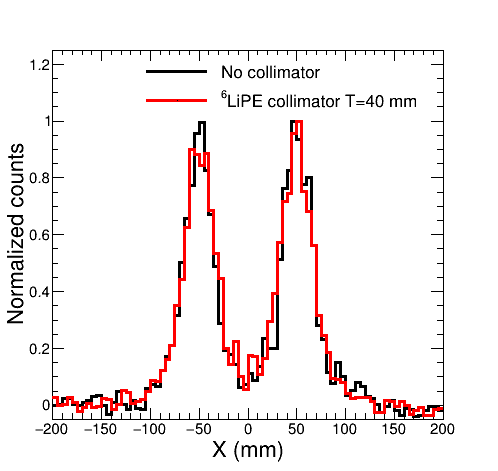}
\end{center}
\caption{Reconstructed Compton images for a 500~keV \g-ray source placed at 200~mm from GN-Vision without a neutron collimator (top left) and with a 40~mm thick \lipe collimator (D = 2.5~mm) (top right). Bottom: Projections along the X axis of the two images showing the impact of the neutron collimator.}
\label{fig:ComptonImageCollimator}
\end{figure}

The \g-ray imaging performance of GN-Vision and the impact of the mechanical neutron collimator has been studied on the basis of Compton images reconstruced from the simulated data. Figure~\ref{fig:ComptonImageCollimator} shows the \g-ray images obtained with GN-Vision for two point-like sources of 500~keV \g-rays placed at 200~mm of the device and at $\pm$50~mm from the detector axis. The analytical algorithm of Tomitani et al., described in Ref.~\cite{Lerendegui:22}, has been used to reconstruct these images. The two Compton images correspond to simulations of GN-Vision without the \lipe neutron collimator and with the thickest pin-hole collimator (T=40~mm) considered in Sec.~\ref{sec:ResultsNeutrons}.  

The projections of the two Compton images are displayed in the bottom panel of Fig.~\ref{fig:ComptonImageCollimator}. From this figure, one concludes that the good spatial resolution (FWHM) of 33~mm (FWHM), which corresponds to an angular resolution of 9$^\circ$, is not affected by the presence of the neutron pin-hole collimator even for a \g-ray energy as low as 500~keV. A crucial factor for the attained resolution is the use of an evolved Compton algorithm, which clearly outperforms the simple and fast back-projection method~\cite{Babiano20}, as it was shown in previous works of the predecessor i-TED detector~\cite{Lerendegui:22,Lerendegui:22b}. The spatial resolution of the Compton images is ascribed to the uncertainty in the determination of the Compton angle, and it would be significantly enhanced for higher \g-ray energies due to the better energy resolution~\cite{Babiano20}.  

The efficiency of GN-Vision for the imaging of \g-rays of energies from 200~keV to 5 MeV is summarized in Table~\ref{tab:effGamma}. To compute the values in this table, the absolute efficiency has been calculated from the number of S\&A coincidences per number of emitted \g-rays obtained from the MC simulations of a point-like \g-ray source located at 20~cm from the entrance of the \lipe collimator. A realistic threshold of 100~keV in deposited energy per crystal has been assumed. The intrinsic efficiency has been then computed using the solid angle subtended by the Compton scatter plane of 50$\times$50~mm. 
The impact of the separation between the S- and A-planes, which can be remotely adjusted for an optimum trade-off between efficiency and resolution, has been studied. The latter approach known as electronic-dynamic collimation, was already implemented in i-TED, as it is described in Ref.~\cite{Babiano20}.

\begin{table}[ht]
\centering
\caption{Intrinsic detection efficiency of GN-Vision in S\&A coincidence for \g-rays of energies ranging from 500 keV to 5~MeV. Each column shows the result for a different distance between the S- and A-planes. The uncertainties due to counting statistics are below 0.5\%.}
\label{tab:effGamma}
\begin{tabular}{lccc}
\hline
                        &   \multicolumn{3}{c}{Focal distance (mm)}\\ 
\hline
Energy (MeV)        &  5   &   15  &  30  \\
\hline
0.5         &   1.72$\times$10$^{-2}$     &    1.31$\times$10$^{-2}$    & 8.43$\times$10$^{-3}$\\
1.0          &   2.18$\times$10$^{-2}$     &   1.80$\times$10$^{-2}$     & 1.36$\times$10$^{-2}$\\
2.0          &   2.10$\times$10$^{-2}$     &  1.74$\times$10$^{-2}$      & 1.35$\times$10$^{-2}$\\
5.0          &  2.79$\times$10$^{-2}$      &  2.33$\times$10$^{-2}$      & 1.84$\times$10$^{-2}$\\
\hline
\end{tabular}
\end{table}

 The results presented in this section have shown the capability of GN-Vision for the imaging of \g-rays despite the presence of a neutron collimator. Detection efficiency and image resolution are two of the most relevant performance aspects of any Compton imager. The attainable image resolution with GN-Vision is as good as that reported from MC simulations of recent compact dual imaging devices~\cite{Hamrashdi:20}. Moreover, our device provides a two-fold enhancement in resolution compared to large liquid scintillation arrays, which are limited by the low energy resolution of the detectors~\cite{Poitrasson:15,Poitrasson:14}. 
 
 The detection efficiency has to be high enough to reconstruct real-time images in field measurements and real scenarios. A high intrinsic efficiency for the imaging of \g-rays, in the order of 10$^{-2}$ (see Table~\ref{tab:effGamma}), is obtained with GN-Vision thanks to the use of large monolithic crystals and the extended absorber plane composed of four scintillator crystals. These values clearly out-perform other compact dual imaging systems found in the literature, which report intrinsic efficiencies smaller than 10$^{-3}$ for similar devices based on multiple detection layers~\cite{Hamrashdi:20,Steinberger:20}. The large \g-ray imaging efficiency of GN-Vision device is only overcome by bulky devices based on large arrays of scintillators (see for instance Ref.~\cite{Poitrasson:14}). 
 
\section{GN-Vision in a realistic scenario: A nuclear inspection}\label{sec:NuclearIns}

The simulations presented in the previous sections were aimed at illustrating the dual \g-ray and neutron imaging capability of GN-Vision and also to optimize its design. However, these calculations were based on mono-energetic sources, which are not representative of any real scenario or potential application. To show the imaging capabilities of the proposed device in a realistic case, we have simulated the inspection with GN-Vision of a container hiding a sample of $^{252}$Cf (2.645(8)~y), a natural emitter of both neutrons and \g-rays by means of a Spontaneous Fission (SF) decay. This isotope features a very well known neutron fission spectrum~\cite{Mannhart1987,Green2017}, similar to that emitted by sensible nuclear materials such as Uranium and Plutonium~\cite{Steinberger:20}.
 
The simulated scenario consisted on a 10$\times$10~mm cylindrical sample of $^{252}$Cf hidden in a polyethylene container that could be used in real life to moderate the emitted neutrons, hence avoiding their detection by fast neutron detectors based on organic scintillators~\cite{Pozzi:12,Poitrasson:14}. The container was placed at~50 cm from the imaging device. Different dimensions of the container, ranging from 10$\times$10~cm to 20$\times$20~cm, where simulated to study the impact of the thickness in the outgoing neutron spectrum.

\begin{figure}[htbp!]
\begin{center}
  \includegraphics[width=0.7\columnwidth]{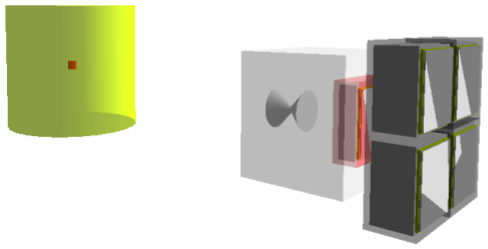}
\caption{Realistic setup implemented in Geant4 to simulate an inspection with GN-vision of a polyethylene container with a hidden $^{252}$Cf source inside (red cylinder).}
\end{center}
\label{fig:Realistic}
\end{figure}

As for the critical parameters of the GN-Vision neutron collimator (see Fig.~\ref{fig:MC}), the same focal distance of F=40~mm used for the conceptual design was chosen. The collimator thickness T=20~mm, that provided good images for 1~eV neutrons was selected (see Fig.~\ref{fig:NeutronImpactD}), and the pinhole aperture was set to 5~mm in order to enhance detection efficiency at a certain cost of image resolution (see Table~\ref{tab:Eff}). The latter is not critical for the imaging of big objects, though. A global view of the simulated setup is shown in Fig.~\ref{fig:Realistic}.

The geometry model of Fig.~\ref{fig:Realistic} was implemented in the \textsc{Geant4} application and the Physics models described in Sec.\ref{sec:MC} were employed. The simulation of the SF decay of $^{252}$Cf is implemented as part of the G4RadioactiveDecay in \textsc{Geant4} 10.6 and later versions~\cite{Geant4106_ReleaseNotes}. Neutrons and \g-rays are emitted according to empirical spectral models, in particular the neutrons use the spectrum of Ref.~\cite{Mannhart1987}.

\begin{figure}[b!]
\begin{center}
  \includegraphics[width=0.45\columnwidth]{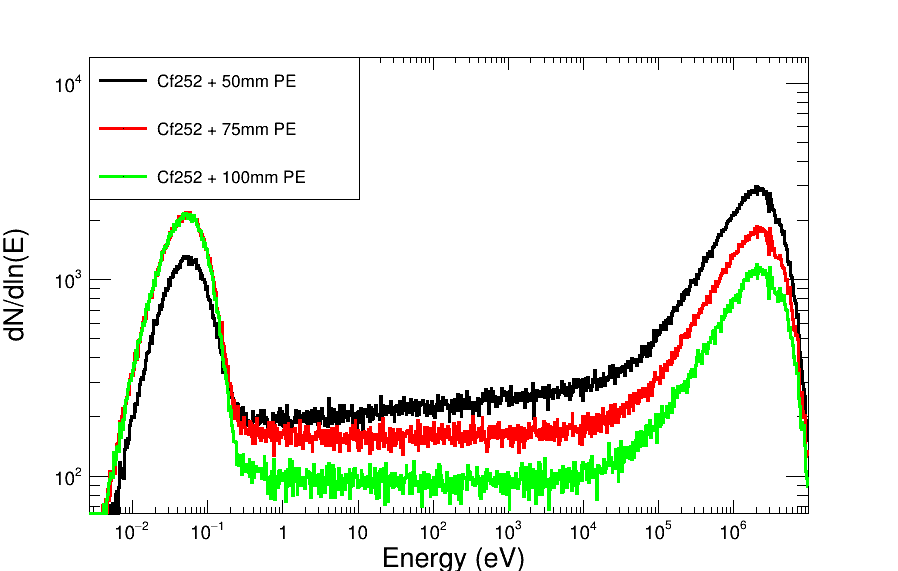}
   \includegraphics[width=0.45\columnwidth]{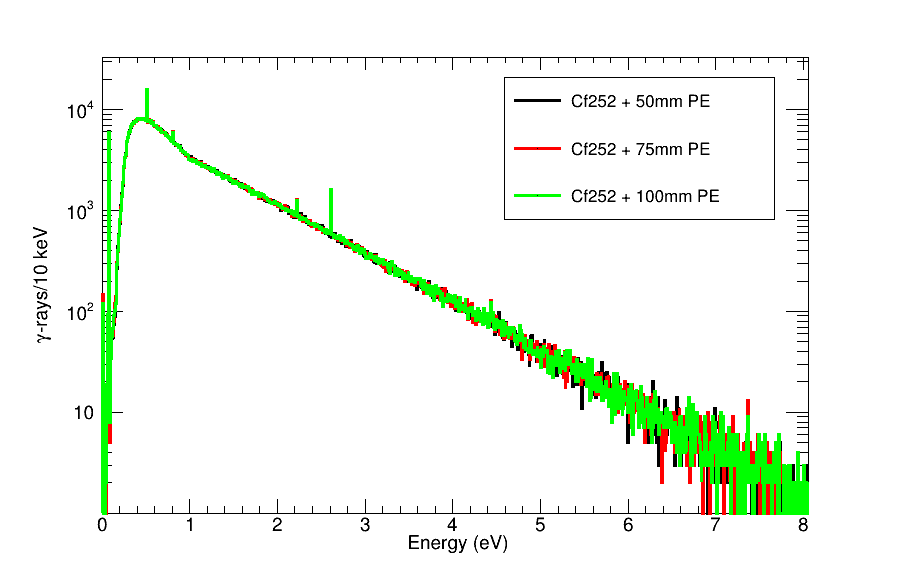}
 \end{center}
\caption{Energy spectra of the $^{252}$Cf SF neutrons after partial moderation in a polyethylene container (left) and \g-rays escaping from the sample (right). The different colours correspond to various dimensions of the container.}
\label{fig:NeutGammaCf252}
\end{figure}

In order to optimize the computing time and avoid the whole simulation of the $^{252}$Cf decay and the transport and moderation of the neutrons in the polyethylene many times, the simulations were carried out in a two-step process. In a first simulation, we simulated the decay of 5$\times$10$^{6}$ $^{252}$Cf nuclei located in a random position within the sample. The outgoing neutrons were transported across the polyethylene container and registered in its outer surface (see Fig.~\ref{fig:Realistic}). The \g-rays were registered in the interface between the $^{252}$Cf sample and the polyethylene. Fig.~\ref{fig:NeutGammaCf252} shows the energy spectra of the registered neutrons and \g-rays. Neutrons are emitted in the SF decay of $^{252}$Cf at a rate of about 1 neutron every 10 decays and energies ranging from 1 to 10 MeV~\cite{Mannhart1987}. After their partial moderation in the polyethylene, the fraction of outgoing neutrons with energies below 100~eV, within the imaging range of GN-Vision, represents 29\% of the total spectrum for a small container of only 10$\times$10~cm. The fraction of slow neutrons increases up to 57\% of the case of the 20$\times$20~cm polyethylene matrix. These results reflect the relevance of being sensitive to the slow neutrons for the imaging of neutron sources hidden in H-rich materials.

The rate of SF \g-rays is about 1.5 every 10 decays. Their energy spectra, shown in the right panel of Fig. ~\ref{fig:NeutGammaCf252}, expands up to 9~MeV, with a maximum below 1~MeV where the Compton imaging technique is proven to work reliably (see Sec.~\ref{sec:ResultsGammas}). As expected, the \g-ray spectra do not get affected by the surrounding polyethylene collimator.  

\begin{figure}[t!]
\begin{center}
  \includegraphics[width=0.9\columnwidth]{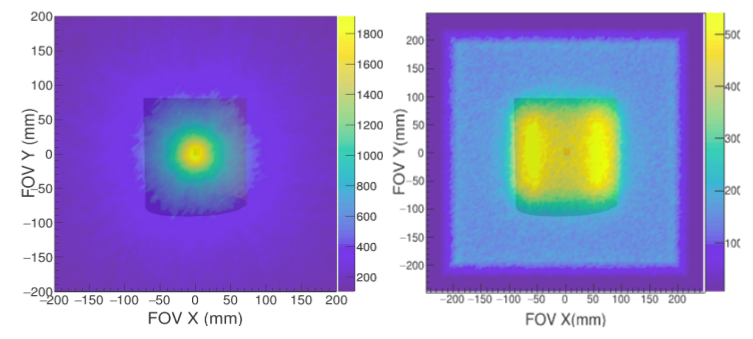}
\caption{\g-ray (left) and neutron (right) images of a 15$\times$15~cm polyethylene container with a hidden $^{252}$Cf source. The shape of the container has been overlapped with the reconstructed images.}
\end{center}
\label{fig:ResultsRealistic}
\end{figure}

The energy distributions of Fig.~\ref{fig:NeutGammaCf252} together with the spatial distributions of the registered particles were re-sampled in the second simulation step, in which a total of 5$\times$10$^{9}$ neutrons and \g-rays were generated aiming at extracting the response of the detector with enough statistics for a proper dual image reconstruction. 
Neutron and \g-ray events for the reconstruction of the images were selected from the output of the simulation as it has been described, respectively, in Secs.~\ref{sec:ResultsGammas} and ~\ref{sec:ResultsNeutrons}. For the case of the \g-rays, S\&A coincidence events with a total energy deposition from 0.5 to 1 MeV were selected. This energy window corresponds to the maximum of the \g-ray spectrum emitted from the $^{252}$Cf sample (see Fig.~\ref{fig:NeutGammaCf252}). SNM such as plutonium have in addition several low energy \g-ray lines (300–400 keV)~\cite{Poitrasson:14}, which would enhance the peak to background ratio of the image.
 
 The simultaneous imaging capability of GN-Vision was evaluated for the three different dimensions of the container. Fig.~\ref{fig:ResultsRealistic} shows the results obtained for the 15$\times$15~cm cylindrical container. For a better illustration of the \g-ray and neutron images, the 2D histograms of the reconstructed figures have been overlapped with the actual geometry of the container. As one can see in Fig.~\ref{fig:ResultsRealistic}, the \g-ray image allows a precise localization of the emitting source. Moreover, the neutron image still preserves the information of the source, albeit thermalized and extended with respect to the true original source, leading to complementary information about the dimensions of the neutron moderator encapsulating the SNM source.
 
 The results presented herein serve to validate the performance of GN-Vision for the identification of SNM hidden in a neutron moderator by means of the simultaneous imaging of both \g-rays and neutrons. 

\section{Summary and outlook}\label{sec:Summary}

In this work we have presented GN-Vision, a novel imaging system which is able to simultaneously detect and localize sources of low energy neutrons and \g-rays. The dual imaging capability is achieved in a single, compact and lightweight device. The latter properties make it very well suited for nuclear safety and control and in nuclear security inspections, where sensible materials naturally emit both neutrons and \g-rays.

The GN-Vision concept consists of two planes of position sensitive detectors, LaCl$_{3}$ and CLYC-6, which exploits the Compton technique for \g-ray imaging.  A mechanical collimator attached to the first plane enables the imaging of slow neutrons ($<$~100 eV). A first implementation of the GN-Vision prototype is based on the early i-TED Compton imager.

This manuscript has presented the conceptual design and optimization of the proposed dual imaging device on the basis of accurate Monte Carlo simulations. The latter have shown that a simple pin-hole geometry for the neutron collimator with a thickness of 20-40~mm and a pinhole aperture of 1-5~mm is able of generating images for neutron energies ranging from thermal to 100 eV while not affecting the Compton imaging of \g-rays in a sizable manner. The specific collimator geometry can be adapted to the final application, for a trade-off between neutron efficiency and image resolution. Last, we have shown the potential applicability of such device for the identification of SNM emitting both neutrons and \g-rays hidden in a neutron-moderating container.

The proposed device has been recently patented~\cite{Patent} and it is currently under development. While the \g-imaging capability is already at a very high technology readiness level (TRL) of 6 following the developments of the previous i-TED Compton imager, the neutron imaging capability has just been conceptually proven for the first time in this work. The experimental integration of the neutron-gamma discrimination with the CLYC-6 crystal utilizing the compact PETSys electronics is currently undergoing the first experimental tests. The full development of GN-Vision and first field test-measurements will require further R\&D that will follow in the upcoming years.

\section*{Declarations}
\begin{backmatter}
\section*{Availability of data and materials}
No experimental data are included in this publication. The code used for the simulations during the current study are available from the corresponding author on reasonable request.

\section*{Competing interests}
The authors declare that they have no known competing financial interests or personal relationships that could have appeared to influence the work reported in this paper.

\section*{Funding}
This work is a development that arises from the i-TED detection system that has been designed and constructed in the framework of the ERC Grant Agreement Nr. 681740. We acknowledge funding from the Universitat de Val\`encia through the Valoritza i Transfereix Programme. J.~Lerendegui-Marco and J.~Balibrea were supported, respectively, by grants FJC2020-044688-I and ICJ220-045122-I funded by MCIN/AEI/ 10.13039/501100-011033 and by European Union NextGenerationEU/PRTR.

\section*{Authors' contributions}
\textbf{J. Lerendegui-Marco:} Investigation, Methodology, Formal analysis, Data curation, Visualization, Software, Writing - original draft, Funding acquisition.
\textbf{V.~Babiano:} Investigation, Software. 
\textbf{J. Balibrea-Correa:} Investigation, Formal analysis, Software, Writing -review.
\textbf{L.~Caballero:} Investigation , Writing -review.
\textbf{D.~Calvo:} Investigation.
\textbf{I. Ladarescu:} Methodology, Software.  \textbf{C. Domingo-Pardo:} Conceptualization, Investigation, Methodology, Supervision, Writing -review \& editing, Project administration, Funding acquisition. 

\section*{Acknowledgements}
 We acknowledge the support of the Agencia Valenciana de Innovaci\'on (AVI) through the Scientific Unit for Business Innovation (Unidad Científica de Innovación Empresarial - UCIE) of IFIC and a collaborative effort with the Instituto Tecnol\'ogico del Pl\'astico (AIMPLAS). We would like to thank J.V. Civera, D. Tchogna of the Mechanics Workshop at IFIC for their excellent CAD design and manufacturing work towards the first prototype.



\bibliographystyle{bmc-mathphys} 

\newcommand{\BMCxmlcomment}[1]{}

\BMCxmlcomment{

<refgrp>

<bibl id="B1">
  <title><p>The use of ionising radiation to image nuclear fuel: A
  review</p></title>
  <aug>
    <au><snm>Parker</snm><fnm>HMO</fnm></au>
    <au><snm>Joyce</snm><fnm>MJ</fnm></au>
  </aug>
  <source>Progress in Nuclear Energy</source>
  <pubdate>2015</pubdate>
  <volume>85</volume>
  <fpage>297</fpage>
  <lpage>318</lpage>
  <url>https://www.sciencedirect.com/science/article/pii/S0149197015300111</url>
</bibl>

<bibl id="B2">
  <title><p>Angular-resolution and material-characterization measurements for a
  dual-particle imaging system with mixed-oxide fuel</p></title>
  <aug>
    <au><snm>Poitrasson Rivière</snm><fnm>A</fnm></au>
    <au><snm>Polack</snm><fnm>JK</fnm></au>
    <au><snm>Hamel</snm><fnm>MC</fnm></au>
    <au><snm>Klemm</snm><fnm>DD</fnm></au>
    <au><snm>Ito</snm><fnm>K</fnm></au>
    <au><snm>McSpaden</snm><fnm>AT</fnm></au>
    <au><snm>Flaska</snm><fnm>M</fnm></au>
    <au><snm>Clarke</snm><fnm>SD</fnm></au>
    <au><snm>Pozzi</snm><fnm>SA</fnm></au>
    <au><snm>Tomanin</snm><fnm>A</fnm></au>
    <au><snm>Peerani</snm><fnm>P</fnm></au>
  </aug>
  <source>Nuclear Instruments and Methods in Physics Research Section A:
  Accelerators, Spectrometers, Detectors and Associated Equipment</source>
  <pubdate>2015</pubdate>
  <volume>797</volume>
  <fpage>278</fpage>
  <lpage>284</lpage>
  <url>https://www.sciencedirect.com/science/article/pii/S0168900215007950</url>
</bibl>

<bibl id="B3">
  <title><p>Rapid imaging of special nuclear materials for nuclear
  nonproliferation and terrorism prevention</p></title>
  <aug>
    <au><snm>Petrović</snm><fnm>J</fnm></au>
    <au><snm>Göök</snm><fnm>A</fnm></au>
    <au><snm>Cederwall</snm><fnm>B</fnm></au>
  </aug>
  <source>Science Advances</source>
  <pubdate>2021</pubdate>
  <volume>7</volume>
  <issue>21</issue>
  <fpage>eabg3032</fpage>
  <url>https://www.science.org/doi/abs/10.1126/sciadv.abg3032</url>
</bibl>

<bibl id="B4">
  <title><p>Radiation imaging using a compact Compton camera mounted on a
  crawler robot inside reactor buildings of Fukushima Daiichi Nuclear Power
  Station</p></title>
  <aug>
    <au><snm>Sato</snm><fnm>Y</fnm></au>
    <au><snm>Terasaka</snm><fnm>Y</fnm></au>
    <au><snm>Utsugi</snm><fnm>W</fnm></au>
    <au><snm>Kikuchi</snm><fnm>H</fnm></au>
    <au><snm>Kiyooka</snm><fnm>H</fnm></au>
    <au><snm>Torii</snm><fnm>T</fnm></au>
  </aug>
  <source>Journal of Nuclear Science and Technology</source>
  <publisher>Taylor & Francis</publisher>
  <pubdate>2019</pubdate>
  <volume>56</volume>
  <issue>9-10</issue>
  <fpage>801</fpage>
  <lpage>808</lpage>
  <url>https://doi.org/10.1080/00223131.2019.1581111</url>
</bibl>

<bibl id="B5">
  <title><p>Gamma-Ray imaging for nuclear security and safety: Towards 3-D
  gamma-ray vision</p></title>
  <aug>
    <au><snm>Vetter</snm><fnm>K</fnm></au>
    <au><snm>Barnowksi</snm><fnm>R</fnm></au>
    <au><snm>Haefner</snm><fnm>A</fnm></au>
    <au><snm>Joshi</snm><fnm>TH</fnm></au>
    <au><snm>Pavlovsky</snm><fnm>R</fnm></au>
    <au><snm>Quiter</snm><fnm>BJ</fnm></au>
  </aug>
  <source>Nuclear Instruments and Methods in Physics Research Section A:
  Accelerators, Spectrometers, Detectors and Associated Equipment</source>
  <pubdate>2018</pubdate>
  <volume>878</volume>
  <fpage>159</fpage>
  <lpage>168</lpage>
  <url>https://www.sciencedirect.com/science/article/pii/S0168900217309269</url>
  <note>Radiation Imaging Techniques and Applications</note>
</bibl>

<bibl id="B6">
  <title><p>First demonstration of aerial gamma-ray imaging using drone for
  prompt radiation survey in Fukushima</p></title>
  <aug>
    <au><snm>Mochizuki</snm><fnm>S.</fnm></au>
    <au><snm>Kataoka</snm><fnm>J.</fnm></au>
    <au><snm>Tagawa</snm><fnm>L.</fnm></au>
    <au><snm>Iwamoto</snm><fnm>Y.</fnm></au>
    <au><snm>Okochi</snm><fnm>H.</fnm></au>
    <au><snm>Katsumi</snm><fnm>N.</fnm></au>
    <au><snm>Kinno</snm><fnm>S.</fnm></au>
    <au><snm>Arimoto</snm><fnm>M.</fnm></au>
    <au><snm>Maruhashi</snm><fnm>T.</fnm></au>
    <au><snm>Fujieda</snm><fnm>K.</fnm></au>
    <au><snm>Kurihara</snm><fnm>T.</fnm></au>
    <au><snm>Ohsuka</snm><fnm>S.</fnm></au>
  </aug>
  <source>Journal of Instrumentation</source>
  <pubdate>2017</pubdate>
  <volume>12</volume>
  <issue>11</issue>
  <fpage>P11014</fpage>
  <url>https://dx.doi.org/10.1088/1748-0221/12/11/P11014</url>
</bibl>

<bibl id="B7">
  <title><p>DUAL PARTICLE IMAGING SYSTEM FOR STANDOFF SNM DETECTION IN
  HIGH-BACKGROUND RADIATION ENVIRONMENT</p></title>
  <aug>
    <au><snm>al.</snm><fnm>SP</fnm></au>
  </aug>
  <pubdate>US2012/0256094A1 Oct. 2012, Priority: 6.4.2012</pubdate>
</bibl>

<bibl id="B8">
  <title><p>{An imaging neutron/gamma-ray spectrometer}</p></title>
  <aug>
    <au><snm>Madden</snm><fnm>AC</fnm></au>
    <au><snm>Bloser</snm><fnm>PF</fnm></au>
    <au><snm>Fourguette</snm><fnm>D</fnm></au>
    <au><snm>Larocque</snm><fnm>L</fnm></au>
    <au><snm>Legere</snm><fnm>JS</fnm></au>
    <au><snm>Lewis</snm><fnm>M</fnm></au>
    <au><snm>McConnell</snm><fnm>ML</fnm></au>
    <au><snm>Rousseau</snm><fnm>M</fnm></au>
    <au><snm>Ryan</snm><fnm>JM</fnm></au>
  </aug>
  <source>Chemical, Biological, Radiological, Nuclear, and Explosives (CBRNE)
  Sensing XIV</source>
  <publisher>SPIE</publisher>
  <editor>Augustus Way Fountain III</editor>
  <pubdate>2013</pubdate>
  <volume>8710</volume>
  <fpage>87101L</fpage>
  <url>https://doi.org/10.1117/12.2018075</url>
</bibl>

<bibl id="B9">
  <title><p>Dual-particle imaging system based on simultaneous detection of
  photon and neutron collision events</p></title>
  <aug>
    <au><snm>Poitrasson Rivière</snm><fnm>A</fnm></au>
    <au><snm>Hamel</snm><fnm>MC</fnm></au>
    <au><snm>Polack</snm><fnm>JK</fnm></au>
    <au><snm>Flaska</snm><fnm>M</fnm></au>
    <au><snm>Clarke</snm><fnm>SD</fnm></au>
    <au><snm>Pozzi</snm><fnm>SA</fnm></au>
  </aug>
  <source>Nuclear Instruments and Methods in Physics Research Section A:
  Accelerators, Spectrometers, Detectors and Associated Equipment</source>
  <pubdate>2014</pubdate>
  <volume>760</volume>
  <fpage>40</fpage>
  <lpage>45</lpage>
  <url>https://www.sciencedirect.com/science/article/pii/S0168900214005889</url>
</bibl>

<bibl id="B10">
  <title><p>A fast and portable imager for neutron and gamma emitting
  radionuclides</p></title>
  <aug>
    <au><snm>{Al Hamrashdi}</snm><fnm>H</fnm></au>
    <au><snm>Cheneler</snm><fnm>D</fnm></au>
    <au><snm>Monk</snm><fnm>SD</fnm></au>
  </aug>
  <source>Nuclear Instruments and Methods in Physics Research Section A:
  Accelerators, Spectrometers, Detectors and Associated Equipment</source>
  <pubdate>2020</pubdate>
  <volume>953</volume>
  <fpage>163253</fpage>
  <url>https://www.sciencedirect.com/science/article/pii/S0168900219315256</url>
</bibl>

<bibl id="B11">
  <title><p>Imaging Special Nuclear Material using a Handheld Dual Particle
  Imager</p></title>
  <aug>
    <au><snm>Steinberger</snm><fnm>WM</fnm></au>
    <au><snm>Ruch</snm><fnm>ML</fnm></au>
    <au><snm>Giha</snm><fnm>N</fnm></au>
    <au><snm>Fulvio</snm><fnm>AD</fnm></au>
    <au><snm>Marleau</snm><fnm>P</fnm></au>
    <au><snm>Clarke</snm><fnm>SD</fnm></au>
    <au><snm>Pozzi</snm><fnm>SA</fnm></au>
  </aug>
  <source>Scientific Reports</source>
  <pubdate>2020</pubdate>
  <volume>10</volume>
  <issue>1</issue>
  <fpage>1855</fpage>
  <url>https://doi.org/10.1038/s41598-020-58857-z</url>
</bibl>

<bibl id="B12">
  <title><p>The Impact of Neutrons in Clinical Proton Therapy</p></title>
  <aug>
    <au><snm>Schneider</snm><fnm>U</fnm></au>
    <au><snm>Hälg</snm><fnm>R</fnm></au>
  </aug>
  <source>Frontiers in Oncology</source>
  <pubdate>2015</pubdate>
  <volume>5</volume>
  <url>https://www.frontiersin.org/articles/10.3389/fonc.2015.00235</url>
</bibl>

<bibl id="B13">
  <title><p>Proton beam therapy in Europe: more centres need more
  research</p></title>
  <aug>
    <au><snm>Durante</snm><fnm>M</fnm></au>
  </aug>
  <source>British Journal of Cancer</source>
  <pubdate>2019</pubdate>
  <volume>120</volume>
  <issue>8</issue>
  <fpage>777</fpage>
  <lpage>778</lpage>
  <url>https://doi.org/10.1038/s41416-018-0329-x</url>
</bibl>

<bibl id="B14">
  <title><p>A scintillator-based approach to monitor secondary neutron
  production during proton therapy</p></title>
  <aug>
    <au><snm>Clarke</snm><fnm>S. D.</fnm></au>
    <au><snm>Pryser</snm><fnm>E.</fnm></au>
    <au><snm>Wieger</snm><fnm>B. M.</fnm></au>
    <au><snm>Pozzi</snm><fnm>S. A.</fnm></au>
    <au><snm>Haelg</snm><fnm>R. A.</fnm></au>
    <au><snm>Bashkirov</snm><fnm>V. A.</fnm></au>
    <au><snm>Schulte</snm><fnm>R. W.</fnm></au>
  </aug>
  <source>Medical Physics</source>
  <pubdate>2016</pubdate>
  <volume>43</volume>
  <issue>11</issue>
  <fpage>5915</fpage>
  <lpage>5924</lpage>
  <url>https://aapm.onlinelibrary.wiley.com/doi/abs/10.1118/1.4963813</url>
</bibl>

<bibl id="B15">
  <title><p>Device for the simultaneous Detection, Identification,
  Quantification and/or location of gamma radiation and neutron
  sources</p></title>
  <aug>
    <au><snm>Lerendegui Marco</snm><fnm>J.</fnm></au>
    <au><snm>Balibrea Correa</snm><fnm>J.</fnm></au>
    <au><snm>Domingo Pardo</snm><fnm>C.</fnm></au>
    <au><snm>Caballero</snm><fnm>L.</fnm></au>
    <au><snm>Babiano</snm><fnm>V.</fnm></au>
    <au><snm>Ladarescu</snm><fnm>I.</fnm></au>
  </aug>
  <pubdate>2021</pubdate>
  <note>International Patent WO2021229132A1</note>
</bibl>

<bibl id="B16">
  <title><p>{i-TED: A novel concept for high-sensitivity (n,{$\gamma$})
  cross-section measurements}</p></title>
  <aug>
    <au><snm>{Domingo-Pardo}</snm><fnm>C.</fnm></au>
  </aug>
  <source>Nuclear Instruments and Methods in Physics Research A</source>
  <pubdate>2016</pubdate>
  <volume>825</volume>
  <fpage>78</fpage>
  <lpage>86</lpage>
</bibl>

<bibl id="B17">
  <title><p>{First i-TED demonstrator: A Compton imager with Dynamic Electronic
  Collimation}</p></title>
  <aug>
    <au><snm>{Babiano}</snm><fnm>V.</fnm></au>
    <au><snm>{Balibrea}</snm><fnm>J.</fnm></au>
    <au><snm>{Caballero}</snm><fnm>L.</fnm></au>
    <au><snm>{Calvo}</snm><fnm>D.</fnm></au>
    <au><snm>{Ladarescu}</snm><fnm>I.</fnm></au>
    <au><snm>{Lerendegui}</snm><fnm>J.</fnm></au>
    <au><snm>{Mira Prats}</snm><fnm>S.</fnm></au>
    <au><snm>{Domingo-Pardo}</snm><fnm>C.</fnm></au>
  </aug>
  <source>Nuclear Instruments and Methods in Physics Research A</source>
  <pubdate>2020</pubdate>
  <volume>953</volume>
  <fpage>163228</fpage>
</bibl>

<bibl id="B18">
  <title><p>{Imaging neutron capture cross sections: i-TED proof-of-concept and
  future prospects based on Machine-Learning techniques}</p></title>
  <aug>
    <au><snm>{Babiano}</snm><fnm>V.</fnm></au>
    <au><snm>{Lerendegui-Marco}</snm><fnm>J.</fnm></au>
  </aug>
  <source>{The European Physical Journal A (accepted)}</source>
  <pubdate>2021</pubdate>
  <url>https://arxiv.org/abs/2012.10374</url>
</bibl>

<bibl id="B19">
  <title><p>Towards machine learning aided real-time range imaging in proton
  therapy</p></title>
  <aug>
    <au><snm>Lerendegui Marco</snm><fnm>J</fnm></au>
    <au><snm>Balibrea Correa</snm><fnm>J</fnm></au>
    <au><snm>Babiano Su{\'a}rez</snm><fnm>V</fnm></au>
    <au><snm>Ladarescu</snm><fnm>I</fnm></au>
    <au><snm>Domingo Pardo</snm><fnm>C</fnm></au>
  </aug>
  <source>Scientific Reports</source>
  <pubdate>2022</pubdate>
  <volume>12</volume>
  <issue>1</issue>
  <fpage>2735</fpage>
  <url>https://doi.org/10.1038/s41598-022-06126-6</url>
</bibl>

<bibl id="B20">
  <title><p>{High-sensitivitY Measurements of key stellar Nucleo-Synthesis
  reactions (HYMNS)}</p></title>
  <source>{ERC Consolidator Grant, Agreement No. 681740, PI C.
  Domingo-Pardo}</source>
  <url>https://hymnserc.ific.uv.es</url>
</bibl>

<bibl id="B21">
  <title><p>Passive Gamma-Ray and Neutron Imaging Systems for National Security
  and Nuclear Non-Proliferation in Controlled and Uncontrolled Detection Areas:
  Review of Past and Current Status</p></title>
  <aug>
    <au><snm>Al Hamrashdi</snm><fnm>H</fnm></au>
    <au><snm>Monk</snm><fnm>SD</fnm></au>
    <au><snm>Cheneler</snm><fnm>D</fnm></au>
  </aug>
  <source>Sensors</source>
  <pubdate>2019</pubdate>
  <volume>19</volume>
  <issue>11</issue>
  <url>https://www.mdpi.com/1424-8220/19/11/2638</url>
</bibl>

<bibl id="B22">
  <title><p>Analysis of the BC501A neutron detector signals using the true
  pulse shape</p></title>
  <aug>
    <au><snm>Guerrero</snm><fnm>C.</fnm></au>
    <au><snm>Cano Ott</snm><fnm>D.</fnm></au>
    <au><snm>Fernández Ordóñez</snm><fnm>M.</fnm></au>
    <au><snm>González Romero</snm><fnm>E.</fnm></au>
    <au><snm>Martínez</snm><fnm>T.</fnm></au>
    <au><snm>Villamarín</snm><fnm>D.</fnm></au>
  </aug>
  <source>Nuclear Instruments and Methods in Physics Research Section A:
  Accelerators, Spectrometers, Detectors and Associated Equipment</source>
  <pubdate>2008</pubdate>
  <volume>597</volume>
  <issue>2</issue>
  <fpage>212</fpage>
  <lpage>218</lpage>
  <url>https://www.sciencedirect.com/science/article/pii/S0168900208014113</url>
</bibl>

<bibl id="B23">
  <title><p>The CLYC-6 and CLYC-7 response to $\gamma$-rays, fast and thermal
  neutrons</p></title>
  <aug>
    <au><snm>Giaz</snm><fnm>A.</fnm></au>
    <au><snm>Pellegri</snm><fnm>L.</fnm></au>
    <au><snm>Camera</snm><fnm>F.</fnm></au>
    <au><snm>Blasi</snm><fnm>N.</fnm></au>
    <au><snm>Brambilla</snm><fnm>S.</fnm></au>
    <au><snm>Ceruti</snm><fnm>S.</fnm></au>
    <au><snm>Million</snm><fnm>B.</fnm></au>
    <au><snm>Riboldi</snm><fnm>S.</fnm></au>
    <au><snm>Cazzaniga</snm><fnm>C.</fnm></au>
    <au><snm>Gorini</snm><fnm>G.</fnm></au>
    <au><snm>Nocente</snm><fnm>M.</fnm></au>
    <au><snm>Pietropaolo</snm><fnm>A.</fnm></au>
    <au><snm>Pillon</snm><fnm>M.</fnm></au>
    <au><snm>Rebai</snm><fnm>M.</fnm></au>
    <au><snm>Tardocchi</snm><fnm>M.</fnm></au>
  </aug>
  <source>Nuclear Instruments and Methods in Physics Research Section A:
  Accelerators, Spectrometers, Detectors and Associated Equipment</source>
  <pubdate>2016</pubdate>
  <volume>810</volume>
  <fpage>132</fpage>
  <lpage>139</lpage>
  <url>https://www.sciencedirect.com/science/article/pii/S0168900215015065</url>
</bibl>

<bibl id="B24">
  <title><p>Fast neutron detection efficiency of 6Li and 7Li enriched {CLYC}
  scintillators using an Am-Be source</p></title>
  <aug>
    <au><snm>Blasi</snm><fnm>N.</fnm></au>
    <au><snm>Brambilla</snm><fnm>S.</fnm></au>
    <au><snm>Camera</snm><fnm>F.</fnm></au>
    <au><snm>Ceruti</snm><fnm>S.</fnm></au>
    <au><snm>Giaz</snm><fnm>A.</fnm></au>
    <au><snm>Gini</snm><fnm>L.</fnm></au>
    <au><snm>Groppi</snm><fnm>F.</fnm></au>
    <au><snm>Manenti</snm><fnm>S.</fnm></au>
    <au><snm>Mentana</snm><fnm>A.</fnm></au>
    <au><snm>Million</snm><fnm>B.</fnm></au>
    <au><snm>Riboldi</snm><fnm>S.</fnm></au>
  </aug>
  <source>Journal of Instrumentation</source>
  <publisher>{IOP} Publishing</publisher>
  <pubdate>2018</pubdate>
  <volume>13</volume>
  <issue>11</issue>
  <fpage>P11010</fpage>
  <lpage>-P11010</lpage>
  <url>https://doi.org/10.1088/1748-0221/13/11/p11010</url>
</bibl>

<bibl id="B25">
  <title><p>Gamma-radiation imaging system based on the Compton
  effect</p></title>
  <aug>
    <au><snm>Everett</snm><fnm>D.B.</fnm></au>
    <au><snm>Fleming</snm><fnm>J.S.</fnm></au>
    <au><snm>Todd</snm><fnm>R.W.</fnm></au>
    <au><snm>Nightingale</snm><fnm>J.M.</fnm></au>
  </aug>
  <source>Proceedings of the Institution of Electrical Engineers</source>
  <pubdate>1977</pubdate>
  <volume>124</volume>
  <fpage>995</fpage>
  <lpage>1000(5)</lpage>
  <url>https://digital-library.theiet.org/content/journals/10.1049/piee.1977.0203</url>
</bibl>

<bibl id="B26">
  <title><p>A proposed $\gamma$ camera</p></title>
  <aug>
    <au><snm>TODD</snm><fnm>R. W.</fnm></au>
    <au><snm>NIGHTINGALE</snm><fnm>J. M.</fnm></au>
    <au><snm>EVERETT</snm><fnm>D. B.</fnm></au>
  </aug>
  <source>Nature</source>
  <pubdate>1974</pubdate>
  <volume>251</volume>
  <issue>5471</issue>
  <fpage>132</fpage>
  <lpage>134</lpage>
</bibl>

<bibl id="B27">
  <title><p>A telescope for soft gamma ray astronomy</p></title>
  <aug>
    <au><snm>Schönfelder</snm><fnm>V.</fnm></au>
    <au><snm>Hirner</snm><fnm>A.</fnm></au>
    <au><snm>Schneider</snm><fnm>K.</fnm></au>
  </aug>
  <source>Nuclear Instruments and Methods</source>
  <pubdate>1973</pubdate>
  <volume>107</volume>
  <issue>2</issue>
  <fpage>385</fpage>
  <lpage>394</lpage>
  <url>https://www.sciencedirect.com/science/article/pii/0029554X73902577</url>
</bibl>

<bibl id="B28">
  <title><p>Scintillation Camera</p></title>
  <aug>
    <au><snm>Anger</snm><fnm>HO</fnm></au>
  </aug>
  <source>Review of Scientific Instruments</source>
  <pubdate>1958</pubdate>
  <volume>29</volume>
  <issue>27</issue>
</bibl>

<bibl id="B29">
  <title><p>Experimental Comparison of Knife-Edge and Multi-Parallel Slit
  Collimators for Prompt Gamma Imaging of Proton Pencil Beams</p></title>
  <aug>
    <au><snm>Smeets</snm><fnm>J</fnm></au>
    <au><snm>Roellinghoff</snm><fnm>F</fnm></au>
    <au><snm>Janssens</snm><fnm>G</fnm></au>
    <au><snm>Perali</snm><fnm>I</fnm></au>
    <au><snm>Celani</snm><fnm>A</fnm></au>
    <au><snm>Fiorini</snm><fnm>C</fnm></au>
    <au><snm>Freud</snm><fnm>N</fnm></au>
    <au><snm>Testa</snm><fnm>E</fnm></au>
    <au><snm>Prieels</snm><fnm>D</fnm></au>
  </aug>
  <source>Frontiers in Oncology</source>
  <pubdate>2016</pubdate>
  <volume>6</volume>
  <fpage>156</fpage>
  <url>https://www.frontiersin.org/article/10.3389/fonc.2016.00156</url>
</bibl>

<bibl id="B30">
  <title><p>Gamma-ray imaging system for real-time measurements in nuclear
  waste characterisation</p></title>
  <aug>
    <au><snm>Caballero</snm><fnm>L.</fnm></au>
    <au><snm>Colomer</snm><fnm>FA</fnm></au>
    <au><snm>Bellot</snm><fnm>AC</fnm></au>
    <au><snm>Domingo Pardo</snm><fnm>C.</fnm></au>
    <au><snm>Nieto</snm><fnm>JL</fnm></au>
    <au><snm>Ros</snm><fnm>JA</fnm></au>
    <au><snm>Contreras</snm><fnm>P.</fnm></au>
    <au><snm>Monserrate</snm><fnm>M.</fnm></au>
    <au><snm>Rodr{\'{\i}}guez</snm><fnm>PO</fnm></au>
    <au><snm>Mag{\'{a}}n</snm><fnm>DP</fnm></au>
  </aug>
  <source>Journal of Instrumentation</source>
  <publisher>{IOP} Publishing</publisher>
  <pubdate>2018</pubdate>
  <volume>13</volume>
  <issue>03</issue>
  <fpage>P03016</fpage>
  <lpage>-P03016</lpage>
  <url>https://doi.org/10.1088/1748-0221/13/03/p03016</url>
</bibl>

<bibl id="B31">
  <title><p>{On the performance of large monolithic {LaCl}3(Ce) crystals
  coupled to pixelated silicon photosensors}</p></title>
  <aug>
    <au><snm>Olleros</snm><fnm>P.</fnm></au>
    <au><snm>Caballero</snm><fnm>L.</fnm></au>
    <au><snm>Domingo Pardo</snm><fnm>C.</fnm></au>
    <au><snm>Babiano</snm><fnm>V.</fnm></au>
    <au><snm>Ladarescu</snm><fnm>I.</fnm></au>
    <au><snm>Calvo</snm><fnm>D.</fnm></au>
    <au><snm>Gramage</snm><fnm>P.</fnm></au>
    <au><snm>Nacher</snm><fnm>E.</fnm></au>
    <au><snm>Tain</snm><fnm>J.L.</fnm></au>
    <au><snm>Tolosa</snm><fnm>A.</fnm></au>
  </aug>
  <source>Journal of Instrumentation</source>
  <publisher>{IOP} Publishing</publisher>
  <pubdate>2018</pubdate>
  <volume>13</volume>
  <issue>03</issue>
  <fpage>P03014</fpage>
  <lpage>-P03014</lpage>
  <url>https://doi.org/10.1088
</bibl>

<bibl id="B32">
  <title><p>{{\ensuremath{\gamma}}-Ray position reconstruction in large
  monolithic LaCl$_{3}$(Ce) crystals with SiPM readout}</p></title>
  <aug>
    <au><snm>{Babiano}</snm><fnm>V.</fnm></au>
    <au><snm>{Caballero}</snm><fnm>L.</fnm></au>
    <au><snm>{Calvo}</snm><fnm>D.</fnm></au>
    <au><snm>{Ladarescu}</snm><fnm>I.</fnm></au>
    <au><snm>{Olleros}</snm><fnm>P.</fnm></au>
    <au><snm>{Domingo-Pardo}</snm><fnm>C.</fnm></au>
  </aug>
  <source>Nuclear Instruments and Methods in Physics Research A</source>
  <pubdate>2019</pubdate>
  <volume>931</volume>
  <fpage>1</fpage>
  <lpage>22</lpage>
</bibl>

<bibl id="B33">
  <title><p>{Machine Learning aided 3D-position reconstruction in large LaCl3
  crystals}</p></title>
  <aug>
    <au><snm>Balibrea Correa</snm><fnm>J.</fnm></au>
    <au><snm>Lerendegui Marco</snm><fnm>J.</fnm></au>
    <au><snm>Babiano Suárez</snm><fnm>V.</fnm></au>
    <au><snm>Caballero</snm><fnm>L.</fnm></au>
    <au><snm>Calvo</snm><fnm>D.</fnm></au>
    <au><snm>Ladarescu</snm><fnm>I.</fnm></au>
    <au><snm>Olleros Rodríguez</snm><fnm>P.</fnm></au>
    <au><snm>Domingo Pardo</snm><fnm>C.</fnm></au>
  </aug>
  <source>Nuclear Instruments and Methods in Physics Research Section A:
  Accelerators, Spectrometers, Detectors and Associated Equipment</source>
  <pubdate>2021</pubdate>
  <volume>1001</volume>
  <fpage>165249</fpage>
  <url>https://www.sciencedirect.com/science/article/pii/S0168900221002333</url>
</bibl>

<bibl id="B34">
  <title><p>Hybrid in-beam PET- and Compton prompt-gamma imaging aimed at
  enhanced proton-range verification</p></title>
  <aug>
    <au><cnm>{Balibrea-Correa, J.}</cnm></au>
    <au><cnm>{Lerendegui-Marco, J.}</cnm></au>
    <au><cnm>{Ladarescu, I.}</cnm></au>
    <au><cnm>{Guerrero, C.}</cnm></au>
    <au><cnm>{Rodr\'{\i}guez-Gonz\'alez, T.}</cnm></au>
    <au><cnm>{Jim\'enez-Ramos, M. C.}</cnm></au>
    <au><cnm>{Fern\'andez-Mart\'{\i}nez, B.}</cnm></au>
    <au><cnm>{Domingo-Pardo, C.}</cnm></au>
  </aug>
  <source>Eur. Phys. J. Plus</source>
  <pubdate>2022</pubdate>
  <volume>137</volume>
  <issue>11</issue>
  <fpage>1258</fpage>
  <url>https://doi.org/10.1140/epjp/s13360-022-03414-y</url>
</bibl>

<bibl id="B35">
  <title><p>{TOFPET 2: A high-performance circuit for PET
  time-of-flight}</p></title>
  <aug>
    <au><snm>{Di Francesco}</snm><fnm>A</fnm></au>
    <au><snm>{Bugalho}</snm><fnm>R</fnm></au>
    <au><snm>{Oliveira}</snm><fnm>L</fnm></au>
    <au><snm>{Rivetti}</snm><fnm>A</fnm></au>
    <au><snm>{Rolo}</snm><fnm>M</fnm></au>
    <au><snm>{Silva}</snm><fnm>JC</fnm></au>
    <au><snm>{Varela}</snm><fnm>J</fnm></au>
  </aug>
  <source>Nuclear Instruments and Methods in Physics Research A</source>
  <pubdate>2016</pubdate>
  <volume>824</volume>
  <fpage>194</fpage>
  <lpage>195</lpage>
</bibl>

<bibl id="B36">
  <title><p>{Recent developments in Geant4}</p></title>
  <aug>
    <au><snm>Allison</snm><fnm>J.</fnm></au>
    <au><snm>Amako</snm><fnm>K.</fnm></au>
    <au><snm>Apostolakis</snm><fnm>J.</fnm></au>
    <au><snm>Arce</snm><fnm>P.</fnm></au>
    <au><snm>Asai</snm><fnm>M.</fnm></au>
    <au><snm>Aso</snm><fnm>T.</fnm></au>
    <au><snm>Bagli</snm><fnm>E.</fnm></au>
    <au><snm>Bagulya</snm><fnm>A.</fnm></au>
    <au><snm>Banerjee</snm><fnm>S.</fnm></au>
    <au><snm>Barrand</snm><fnm>G.</fnm></au>
    <au><snm>Beck</snm><fnm>B.R.</fnm></au>
    <au><snm>Bogdanov</snm><fnm>A.G.</fnm></au>
    <au><snm>Brandt</snm><fnm>D.</fnm></au>
    <au><snm>Brown</snm><fnm>J.M.C.</fnm></au>
    <au><snm>Burkhardt</snm><fnm>H.</fnm></au>
    <au><snm>Canal</snm><fnm>P</fnm></au>
    <au><snm>Cano Ott</snm><fnm>D.</fnm></au>
    <au><snm>Chauvie</snm><fnm>S.</fnm></au>
    <au><snm>Cho</snm><fnm>K.</fnm></au>
    <au><snm>Cirrone</snm><fnm>G.A.P.</fnm></au>
    <au><snm>Cooperman</snm><fnm>G.</fnm></au>
    <au><snm>Cortés Giraldo</snm><fnm>M.A.</fnm></au>
    <au><snm>Cosmo</snm><fnm>G.</fnm></au>
    <au><snm>Cuttone</snm><fnm>G.</fnm></au>
    <au><snm>Depaola</snm><fnm>G.</fnm></au>
    <au><snm>Desorgher</snm><fnm>L.</fnm></au>
    <au><snm>Dong</snm><fnm>X.</fnm></au>
    <au><snm>Dotti</snm><fnm>A.</fnm></au>
    <au><snm>Elvira</snm><fnm>V.D.</fnm></au>
    <au><snm>Folger</snm><fnm>G.</fnm></au>
    <au><snm>Francis</snm><fnm>Z.</fnm></au>
    <au><snm>Galoyan</snm><fnm>A.</fnm></au>
    <au><snm>Garnier</snm><fnm>L.</fnm></au>
    <au><snm>Gayer</snm><fnm>M.</fnm></au>
    <au><snm>Genser</snm><fnm>K.L.</fnm></au>
    <au><snm>Grichine</snm><fnm>V.M.</fnm></au>
    <au><snm>Guatelli</snm><fnm>S.</fnm></au>
    <au><snm>Guèye</snm><fnm>P.</fnm></au>
    <au><snm>Gumplinger</snm><fnm>P.</fnm></au>
    <au><snm>Howard</snm><fnm>A.S.</fnm></au>
    <au><snm>Hřivnáčová</snm><fnm>I.</fnm></au>
    <au><snm>Hwang</snm><fnm>S.</fnm></au>
    <au><snm>Incerti</snm><fnm>S.</fnm></au>
    <au><snm>Ivanchenko</snm><fnm>A.</fnm></au>
    <au><snm>Ivanchenko</snm><fnm>V.N.</fnm></au>
    <au><snm>Jones</snm><fnm>F.W.</fnm></au>
    <au><snm>Jun</snm><fnm>S.Y.</fnm></au>
    <au><snm>Kaitaniemi</snm><fnm>P.</fnm></au>
    <au><snm>Karakatsanis</snm><fnm>N.</fnm></au>
    <au><snm>Karamitros</snm><fnm>M.</fnm></au>
    <au><snm>Kelsey</snm><fnm>M.</fnm></au>
    <au><snm>Kimura</snm><fnm>A.</fnm></au>
    <au><snm>Koi</snm><fnm>T.</fnm></au>
    <au><snm>Kurashige</snm><fnm>H.</fnm></au>
    <au><snm>Lechner</snm><fnm>A.</fnm></au>
    <au><snm>Lee</snm><fnm>S.B.</fnm></au>
    <au><snm>Longo</snm><fnm>F.</fnm></au>
    <au><snm>Maire</snm><fnm>M.</fnm></au>
    <au><snm>Mancusi</snm><fnm>D.</fnm></au>
    <au><snm>Mantero</snm><fnm>A.</fnm></au>
    <au><snm>Mendoza</snm><fnm>E.</fnm></au>
    <au><snm>Morgan</snm><fnm>B.</fnm></au>
    <au><snm>Murakami</snm><fnm>K.</fnm></au>
    <au><snm>Nikitina</snm><fnm>T.</fnm></au>
    <au><snm>Pandola</snm><fnm>L.</fnm></au>
    <au><snm>Paprocki</snm><fnm>P.</fnm></au>
    <au><snm>Perl</snm><fnm>J.</fnm></au>
    <au><snm>Petrović</snm><fnm>I.</fnm></au>
    <au><snm>Pia</snm><fnm>M.G.</fnm></au>
    <au><snm>Pokorski</snm><fnm>W.</fnm></au>
    <au><snm>Quesada</snm><fnm>J.M.</fnm></au>
    <au><snm>Raine</snm><fnm>M.</fnm></au>
    <au><snm>Reis</snm><fnm>M.A.</fnm></au>
    <au><snm>Ribon</snm><fnm>A.</fnm></au>
    <au><snm>Fira]</snm><fnm>AR</fnm></au>
    <au><snm>Romano</snm><fnm>F.</fnm></au>
    <au><snm>Russo</snm><fnm>G.</fnm></au>
    <au><snm>Santin</snm><fnm>G.</fnm></au>
    <au><snm>Sasaki</snm><fnm>T.</fnm></au>
    <au><snm>Sawkey</snm><fnm>D.</fnm></au>
    <au><snm>Shin</snm><fnm>J.I.</fnm></au>
    <au><snm>Strakovsky</snm><fnm>I.I.</fnm></au>
    <au><snm>Taborda</snm><fnm>A.</fnm></au>
    <au><snm>Tanaka</snm><fnm>S.</fnm></au>
    <au><snm>Tomé</snm><fnm>B.</fnm></au>
    <au><snm>Toshito</snm><fnm>T.</fnm></au>
    <au><snm>Tran</snm><fnm>H.N.</fnm></au>
    <au><snm>Truscott</snm><fnm>P.R.</fnm></au>
    <au><snm>Urban</snm><fnm>L.</fnm></au>
    <au><snm>Uzhinsky</snm><fnm>V.</fnm></au>
    <au><snm>Verbeke</snm><fnm>J.M.</fnm></au>
    <au><snm>Verderi</snm><fnm>M.</fnm></au>
    <au><snm>Wendt</snm><fnm>B.L.</fnm></au>
    <au><snm>Wenzel</snm><fnm>H.</fnm></au>
    <au><snm>Wright</snm><fnm>D.H.</fnm></au>
    <au><snm>Wright</snm><fnm>D.M.</fnm></au>
    <au><snm>Yamashita</snm><fnm>T.</fnm></au>
    <au><snm>Yarba</snm><fnm>J.</fnm></au>
    <au><snm>Yoshida</snm><fnm>H.</fnm></au>
  </aug>
  <source>Nuclear Instruments and Methods in Physics Research Section A:
  Accelerators, Spectrometers, Detectors and Associated Equipment</source>
  <pubdate>2016</pubdate>
  <volume>835</volume>
  <fpage>186</fpage>
  <lpage>225</lpage>
  <url>http://www.sciencedirect.com/science/article/pii/S0168900216306957</url>
</bibl>

<bibl id="B37">
  <title><p>Lithium Polyethylene, JCS Nuclear Solutions</p></title>
  <source>\url{https://johncaunt.com/products/lithium-polyethylene/}</source>
</bibl>

<bibl id="B38">
  <title><p>{GEANT4 Reference Physics Lists}</p></title>
  <source>\url{https://geant4.web.cern.ch/node/155}</source>
</bibl>

<bibl id="B39">
  <title><p>{New Standard Evaluated Neutron Cross Section Libraries for the
  GEANT4 Code and First Verification}</p></title>
  <aug>
    <au><snm>{Mendoza}</snm><fnm>E.</fnm></au>
    <au><snm>{Cano-Ott}</snm><fnm>D.</fnm></au>
    <au><snm>{Koi}</snm><fnm>T.</fnm></au>
    <au><snm>{Guerrero}</snm><fnm>C.</fnm></au>
  </aug>
  <source>IEEE Transactions on Nuclear Science</source>
  <pubdate>2014</pubdate>
  <volume>61</volume>
  <issue>4</issue>
  <fpage>2357</fpage>
  <lpage>2364</lpage>
</bibl>

<bibl id="B40">
  <title><p>{The joint evaluated fission and fusion nuclear data library,
  JEFF-3.3}</p></title>
  <aug>
    <au><snm>Plompen</snm><fnm>A. J. M.</fnm></au>
    <au><snm>Cabellos</snm><fnm>O.</fnm></au>
    <au><snm>De Saint Jean</snm><fnm>C.</fnm></au>
    <au><snm>Fleming</snm><fnm>M.</fnm></au>
    <au><snm>Algora</snm><fnm>A.</fnm></au>
    <au><snm>Angelone</snm><fnm>M.</fnm></au>
    <au><snm>Archier</snm><fnm>P.</fnm></au>
    <au><snm>Bauge</snm><fnm>E.</fnm></au>
    <au><snm>Bersillon</snm><fnm>O.</fnm></au>
    <au><snm>Blokhin</snm><fnm>A.</fnm></au>
    <au><snm>Cantargi</snm><fnm>F.</fnm></au>
    <au><snm>Chebboubi</snm><fnm>A.</fnm></au>
    <au><snm>Diez</snm><fnm>C.</fnm></au>
    <au><snm>Duarte</snm><fnm>H.</fnm></au>
    <au><snm>Dupont</snm><fnm>E.</fnm></au>
    <au><snm>Dyrda</snm><fnm>J.</fnm></au>
    <au><snm>Erasmus</snm><fnm>B.</fnm></au>
    <au><snm>Fiorito</snm><fnm>L.</fnm></au>
    <au><snm>Fischer</snm><fnm>U.</fnm></au>
    <au><snm>Flammini</snm><fnm>D.</fnm></au>
    <au><snm>Foligno</snm><fnm>D.</fnm></au>
    <au><snm>Gilbert</snm><fnm>M. R.</fnm></au>
    <au><snm>Granada</snm><fnm>J. R.</fnm></au>
    <au><snm>Haeck</snm><fnm>W.</fnm></au>
    <au><snm>Hambsch</snm><fnm>F. J.</fnm></au>
    <au><snm>Helgesson</snm><fnm>P.</fnm></au>
    <au><snm>Hilaire</snm><fnm>S.</fnm></au>
    <au><snm>Hill</snm><fnm>I.</fnm></au>
    <au><snm>Hursin</snm><fnm>M.</fnm></au>
    <au><snm>Ichou</snm><fnm>R.</fnm></au>
    <au><snm>Jacqmin</snm><fnm>R.</fnm></au>
    <au><snm>Jansky</snm><fnm>B.</fnm></au>
    <au><snm>Jouanne</snm><fnm>C.</fnm></au>
    <au><snm>Kellett</snm><fnm>M. A.</fnm></au>
    <au><snm>Kim</snm><fnm>D. H.</fnm></au>
    <au><snm>Kim</snm><fnm>H. I.</fnm></au>
    <au><snm>Kodeli</snm><fnm>I.</fnm></au>
    <au><snm>Koning</snm><fnm>A. J.</fnm></au>
    <au><snm>Konobeyev</snm><fnm>AY</fnm></au>
    <au><snm>Kopecky</snm><fnm>S.</fnm></au>
    <au><snm>Kos</snm><fnm>B.</fnm></au>
    <au><snm>Kr{\'a}sa</snm><fnm>A.</fnm></au>
    <au><snm>Leal</snm><fnm>L. C.</fnm></au>
    <au><snm>Leclaire</snm><fnm>N.</fnm></au>
    <au><snm>Leconte</snm><fnm>P.</fnm></au>
    <au><snm>Lee</snm><fnm>Y. O.</fnm></au>
    <au><snm>Leeb</snm><fnm>H.</fnm></au>
    <au><snm>Litaize</snm><fnm>O.</fnm></au>
    <au><snm>Majerle</snm><fnm>M.</fnm></au>
    <au><snm>M{\'a}rquez Dami{\'a}n</snm><fnm>J. I.</fnm></au>
    <au><snm>Michel Sendis</snm><fnm>F.</fnm></au>
    <au><snm>Mills</snm><fnm>R. W.</fnm></au>
    <au><snm>Morillon</snm><fnm>B.</fnm></au>
    <au><snm>Nogu{\`e}re</snm><fnm>G.</fnm></au>
    <au><snm>Pecchia</snm><fnm>M.</fnm></au>
    <au><snm>Pelloni</snm><fnm>S.</fnm></au>
    <au><snm>Pereslavtsev</snm><fnm>P.</fnm></au>
    <au><snm>Perry</snm><fnm>R. J.</fnm></au>
    <au><snm>Rochman</snm><fnm>D.</fnm></au>
    <au><snm>R{\"o}hrmoser</snm><fnm>A.</fnm></au>
    <au><snm>Romain</snm><fnm>P.</fnm></au>
    <au><snm>Romojaro</snm><fnm>P.</fnm></au>
    <au><snm>Roubtsov</snm><fnm>D.</fnm></au>
    <au><snm>Sauvan</snm><fnm>P.</fnm></au>
    <au><snm>Schillebeeckx</snm><fnm>P.</fnm></au>
    <au><snm>Schmidt</snm><fnm>K. H.</fnm></au>
    <au><snm>Serot</snm><fnm>O.</fnm></au>
    <au><snm>Simakov</snm><fnm>S.</fnm></au>
    <au><snm>Sirakov</snm><fnm>I.</fnm></au>
    <au><snm>Sj{\"o}strand</snm><fnm>H.</fnm></au>
    <au><snm>Stankovskiy</snm><fnm>A.</fnm></au>
    <au><snm>Sublet</snm><fnm>J. C.</fnm></au>
    <au><snm>Tamagno</snm><fnm>P.</fnm></au>
    <au><snm>Trkov</snm><fnm>A.</fnm></au>
    <au><snm>Marck</snm><fnm>S.</fnm></au>
    <au><snm>{\'A}lvarez Velarde</snm><fnm>F.</fnm></au>
    <au><snm>Villari</snm><fnm>R.</fnm></au>
    <au><snm>Ware</snm><fnm>T. C.</fnm></au>
    <au><snm>Yokoyama</snm><fnm>K.</fnm></au>
    <au><snm>{\v{Z}}erovnik</snm><fnm>G.</fnm></au>
  </aug>
  <source>The European Physical Journal A</source>
  <pubdate>2020</pubdate>
  <volume>56</volume>
  <issue>7</issue>
  <fpage>181</fpage>
  <url>https://doi.org/10.1140/epja/s10050-020-00141-9</url>
</bibl>

<bibl id="B41">
  <title><p>Response functions of Cs2LiYCl6: Ce scintillator to neutron and
  gamma radiation</p></title>
  <aug>
    <au><snm>Machrafi</snm><fnm>R</fnm></au>
    <au><snm>Khan</snm><fnm>N</fnm></au>
    <au><snm>Miller</snm><fnm>A.</fnm></au>
  </aug>
  <source>Radiation Measurements</source>
  <pubdate>2014</pubdate>
  <volume>70</volume>
  <fpage>5</fpage>
  <lpage>10</lpage>
</bibl>

<bibl id="B42">
  <title><p>Coded-aperture imaging systems: Past, present and future
  development – A review</p></title>
  <aug>
    <au><snm>Cieślak</snm><fnm>MJ</fnm></au>
    <au><snm>Gamage</snm><fnm>KA</fnm></au>
    <au><snm>Glover</snm><fnm>R</fnm></au>
  </aug>
  <source>Radiation Measurements</source>
  <pubdate>2016</pubdate>
  <volume>92</volume>
  <fpage>59</fpage>
  <lpage>71</lpage>
  <url>https://www.sciencedirect.com/science/article/pii/S1350448716301524</url>
</bibl>

<bibl id="B43">
  <title><p>i-TED: Compton Imaging and Machine-Learning Techniques for Enhanced
  Sensitivity Neutron Capture Time-of-flight Measurements</p></title>
  <aug>
    <au><snm>Lerendegui Marco</snm><fnm>J.</fnm></au>
    <au><snm>Babiano Suárez</snm><fnm>V.</fnm></au>
    <au><snm>Balibrea Correa</snm><fnm>J.</fnm></au>
    <au><snm>Caballero</snm><fnm>L.</fnm></au>
    <au><snm>Calvo</snm><fnm>D.</fnm></au>
    <au><snm>Domingo Pardo</snm><fnm>C.</fnm></au>
    <au><snm>Ladarescu</snm><fnm>I.</fnm></au>
  </aug>
  <source>2021 IEEE Nuclear Science Symposium and Medical Imaging Conference
  (NSS/MIC)</source>
  <pubdate>2021</pubdate>
  <fpage>1</fpage>
  <lpage>7</lpage>
</bibl>

<bibl id="B44">
  <title><p>Evaluation of the Cf-252 fission neutron spectrum between 0 MeV and
  20 MeV</p></title>
  <aug>
    <au><snm>Mannhart</snm><fnm>W.</fnm></au>
  </aug>
  <publisher>International Atomic Energy Agency (IAEA)</publisher>
  <pubdate>1987</pubdate>
  <fpage>158</fpage>
  <lpage>171</lpage>
  <url>http://inis.iaea.org/search/search.aspx?orig_q=RN:18075519</url>
  <note>IAEA-TECDOC--410</note>
</bibl>

<bibl id="B45">
  <title><p>The Californium-252 Fission Neutron Spectrum from 0.5 to 13
  MeV</p></title>
  <aug>
    <au><snm>Green</snm><fnm>L.</fnm></au>
    <au><snm>Mitchell</snm><fnm>J. A.</fnm></au>
    <au><snm>Steen</snm><fnm>N. M.</fnm></au>
  </aug>
  <source>Nuclear Science and Engineering</source>
  <publisher>Taylor & Francis</publisher>
  <pubdate>1973</pubdate>
  <volume>50</volume>
  <issue>3</issue>
  <fpage>257</fpage>
  <lpage>272</lpage>
  <url>https://doi.org/10.13182/NSE73-A28979</url>
</bibl>

<bibl id="B46">
  <title><p>{Geant4 10.6 Release Notes}</p></title>
  <source>\url{https://geant4-data.web.cern.ch/ReleaseNotes/ReleaseNotes4.10.6.html}</source>
</bibl>

</refgrp>
} 



\end{backmatter}
\end{document}